\documentclass[aps, prb, twocolumn, groupedaddress, floatfix]{revtex4-1}
\usepackage{amsmath, amssymb}
\usepackage{braket}
\usepackage{color}
\usepackage{here}
\usepackage{graphicx}
\usepackage{hyperref}
\begin{document}
\title{Abundant quadrupolar or nematic phases driven by the Heisenberg interactions in a spin-1 dimer system forming a bilayer}
\author{Katsuhiro Tanaka and Chisa Hotta}
\affiliation{Department of Basic Science, University of Tokyo, Meguro, Tokyo 153-8902, Japan}
\date{\today}
\begin{abstract}
        We explore several classes of quadrupolar ordering in a system of 
        antiferromagnetically coupled quantum spin-1 dimers, 
        which are stacked in the triangular lattice geometry forming a bilayer. 
        Low-energy properties of this model is described by an $\mathcal{S}=1$ hard-core bosonic degrees of freedom defined on each dimer-bond, 
        where the singlet and triplet states of the dimerized spins are interpreted as the vacuum and the occupancy of boson, respectively. 
        The number of bosons per dimer and the magnetic and density fluctuations of bosons are 
        controlled by the inter-dimer Heisenberg interactions. 
        In a solid phase where each dimer hosts one boson and the inter-dimer interaction is weak, 
        a conventional spin nematic phase is realized by the pair-fluctuation of bosons. 
        Larger inter-dimer interaction favors Bose Einstein condensates (BEC) carrying quadrupolar moments. 
        Among them, we find one exotic phase where the quadrupoles develop a spatially modulated structure on the top of a uniform BEC, 
        interpreted in the original dimerized spin-1 model as coexistent $p$-type nematic and 120$^\circ$-magnetic correlations. 
        This may explain an intriguing nonmagnetic phase found in Ba$_{3}$ZnRu$_{2}$O$_{9}$. 
\end{abstract}
\maketitle
\section{Introduction}
\label{sec:introduction}
  Nematics, regarded as a sort of liquid crystal in a more general context, 
  now forms a wide range of phases of matter in crystalline solids. 
  The ``electronic nematic state'' was first proposed in a doped Mott insulator as a consequence of the melting of stripes, aiming to understand the origin of high-$T_c$ phase in cuprates~\cite{Kivelson1998_Nature_393_550}. 
  More recent example is the nematicity of electronic wave functions induced by the orbital ordering 
  in iron-based superconductors, possibly dominating the stability of the superconductivity
  ~\cite{Laad2011_PhysRevB_84_054530,Onari2012_PhysRevLett_109_137001,Fernandes2014_natphys_10_97}. 
  When defined on a crystal lattice, the nematics of charges and orbitals, 
  and also of spins are all described by the quadrupolar order parameter representing the symmetry of their wave function. 
\par
  In insulating quantum magnets~\cite{Blume1969_JApplPhys_40_1249,Andreev1984_SovPhysJETP_60_267}, 
  nematic phases appear when the spin moments break their rotational symmetry 
  and form a wave function in the shape of rod- or disk-like director which collectively align in space. 
  The search of spin nematics has been a challenge, since proposals relevant to experiments are provided only in limited numbers of systems; 
  in a layered solid $\,^3$He~\cite{Ishida1997_PhysRevLett_79_3451}, 
  in an artificially designed optical lattices~\cite{Kimura2005_PhysRevLett_94_110403,Forges2014_PhysRevLett_113_200402}, 
  and in a quantum spin-1/2 magnets near the saturation field~\cite{Svistov2011_JETPLett_93_21,Kohama2019_PNAS_116_10686,Skoulatos2019_PhysRevB_100_014405}. 
\par
  There had been some reasons that the quantum spin systems cannot easily become a good platform of spin nematics. 
  In materials, the electrons carry spin-1/2, which is a dipole by itself, 
  and to form a quadrupole which is a rank-2 tensor, we need at least spin-1 with three different $\mathcal{S}^{z}$-levels
  (see Eq.~(\ref{eq:quadrupole_operator})). 
  There are two ways to construct spin-1 from spin-1/2. 
  One is to use the Hund's coupling between spin-1/2 in different orbitals on the same site, 
  which will generate \textit{site}-nematics. 
  The other is to efficiently compose spin-1 from two spin-1/2's on neighboring sites by the interaction, 
  in which case the \textit{bond}-nematic is formed. 
  In the former case, a quantum spin-1 bilinear-biquadratic (BLBQ) model 
  is known to host a spin nematic phase for a large biquadratic interaction, 
  $(\boldsymbol{S}_{i}\cdot \boldsymbol{S}_{j})^{2}$, 
  where $\boldsymbol{S}_{i}$ is a spin-1 operator~\cite{Blume1969_JApplPhys_40_1249,Chen1971_PhysRevLett_27_1383,
  	Papanicolaou1988_NuclPhysB_305_367,Tanaka2001_JPhysAMathGen_34_304,
    Harada2002_PhysRevB_65_052403,Tsunetsugu2006_JPhysSocJpn_75_083701,
    Lauchli2006_PhysRevLett_97_087205,Bhattacharjee2006_PhysRevB_74_092406,
    Lauchli2006_PhysRevB_74_144426,Harada2007_JPhysSocJpn_76_013703,
    Li2007_PhysRevB_75_104420,Tsunetsugu2007_JPhysCondensMatter_19_145248,
    Stoudenmire2009_PhysRevB_79_214436,Toth2012_PhysRevB_85_140403,
    Niesen2017_PhysRevB_95_180404,Niesen2017_SciPostPhys_3_030,
    Niesen2018_PhysRevB_97_245146}. 
  The biquadratic interaction works as a strong quantum fluctuation exchanging the spin-1 pairs, 
  and kills the anti-symmetric (dipolar) component. 
  Since the biquadratic interaction is generally much smaller than the Heisenberg (bilinear) 
  interaction~\cite{Tanaka2018_JPhysSocJpn_87_023702}, 
  this spin nematics is hardly realized in materials. 
  To have the latter bond nematics in a spin-1/2 model~\cite{Andreev1984_SovPhysJETP_60_267}, often a very high magnetic field and a frustration effect are required; 
  In a fully polarized spin-1/2 state, the standard lowest energy excitation is an $S^{z}=-1$ magnon. 
  However, if there are good reasons to suppress the kinetics of magnon, 
  e.g. the frustration effect on a $J_1$--$J_2$ square lattice model or a ring exchange model, 
  the lowest excitation is replaced by the multi-magnons propagating together~\cite{
    Momoi2000_PhysRevB_62_15067,Momoi2005_ProgTheorPhysSuppl_159_72,
    Shannon2006_PhysRevLett_96_027213,Momoi2006_PhysRevLett_97_257204,
    Hikihara2008_JPhysSocJpn_77_014709,
    Hikihara2008_PhysRevB_78_144404,Sudan2009_PhysRevB_80_140402,
    Zhitomirsky2010_EurophysLett_92_37001,Momoi2012_PhysRevLett_108_057206,Wang2018_PhysRevLett_120_247201}. 
  For example, the bounded two-magnons consisting of two spin-1/2's pointing downward form 
  a quadrupole by definition, and condense into a spin nematic phase near the saturation field 
  in spin-1/2 ladders~\cite{Hikihara2008_PhysRevB_78_144404,Sudan2009_PhysRevB_80_140402}. 
  Whereas in practice, it is hard to realize such a high field in experiments. 
\par
  Recently, a double-layered spin-1/2 dimer system is proposed as a platform of spin nematics in a \textit{zero magnetic field}~\cite{Yokoyama2018_PhysRevB_97_180404}. 
  When the spin-1/2's are antiferromagnetically coupled within the dimer, 
  they form a singlet, and the inter-dimer Heisenberg interactions work as chemical potential 
  and dope the $\mathcal{S}=1$ triplets. 
  By the additional spin-1/2-four-body inter-dimer interaction, 
  a biquadratic interaction between these doped $\mathcal{S}=1$'s is generated 
  and two different types of spin nematic phases appear next to the singlet phase. 
  The model based on the ferromagnetically coupled spin-1/2 dimers 
  have $\mathcal{S}=1$ triplet on a dimer-bond by construction, 
  and are also found to host a small window of spin nematic phase, when the second order perturbative 
  inter-dimer exchange coupling becomes relevant~\cite{Hikihara2019_PhysRevB_100_214414}. 
  These works show that the spin nematics of the same type as that of the spin-1 BLBQ model 
  is available in the spin-1/2 dimer systems by introducing the inter-dimer four-body exchanges 
  which kill the anti-symmetric part of the spin-1/2 wave function, like the biquadratic interaction does on spin-1. 
\par
  In the present paper, we replace the spin-1/2 dimer in Ref.~\onlinecite{Yokoyama2018_PhysRevB_97_180404}
  with spin-1 dimer, 
  and deal with the two-dimensionally stacked spin-1 dimer forming a bilayer triangular lattice. 
  We show that there are abundant types of magnetic and nonmagnetic long range ordered phases 
  that are described by the quadrupolar as well as dipolar order parameters defined on a dimer-bond. 
  Considering a strong intra-dimer antiferromagnetic and biquadratic coupling, 
  where maximally $\mathcal{S}=1$ moment ($\mathcal{S}=1$ is the on-bond spin-1 throughout the paper) 
  is generated in each dimer out of two spin-1's, 
  the original dimerized spin-1's degrees of freedom is transformed to an $\mathcal{S}=1$ hard-core boson. 
  The inter-dimer Heisenberg spin exchange interaction is then transformed to the kinetic and pair fluctuation effect
  of these bosons as well as the magnetic exchange between $\mathcal{S}=1$'s. 
  When the former fluctuation effect dominates, the $\mathcal{S}=1$ bosons lose its dipolar moment and 
  condense, partially occupying the dimers as quadrupoles which are called FQ-BEC phases. 
  Besides, for a parameter region of nearly decoupled dimers, 
  we find a spin nematic phase; we call ``spin nematics'' the phase based on the bound pairs of $S=1$, 
  typically found in spin-1 BLBQ model. 
  In a dimer-based system, spin nematics can be realized when the bosons are nearly fully packed on dimers. 
  These quadrupolar phases are classified by the types of low-lying tower-of-state excitations 
  in the energy spectrum that tell us how they break the symmetry. 
  The richness of the phase diagram is possibly because of the large fluctuation of spin-1 moments allowed by the 
  larger spin space compared to spin-1/2's, since the former affords larger entanglement in constructing $\mathcal{S}=1$. 
  \par
  The technical aspect of the present work is that, 
  despite a seeming difficulty in increasing the degrees of freedom of the spin moments from spin-1/2 to 1, 
  one can treat it within a low-energy approximation using the same bosonic model 
  as the spin-1/2 dimer case~\cite{Yokoyama2018_PhysRevB_97_180404}. 
  Aside from the transformation of the spin-1/2 model being exact, 
  the only difference is how to map the interaction parameters of spin models to those of the bosonic model. 
  The direct motivation of dealing with spin-1 dimers instead of spin-1/2 dimers is to explain 
  the intriguing nonmagnetic phase of Ba$_{3}M$Ru$_{2}$O$_{9}$ ($M=$ Zn, Ca, etc.)~\cite{Terasaki2017_JPhysSocJpn_86_033702,Yamamoto2018_JPhysCondensMatter_30_355801}, which will be finally discussed. 
\section{Model Hamiltonian}
  We consider a system consisting of dimers of two spin-1's.
  As shown in Fig.~\ref{fig:lattice}(a), 
  the dimers stack parallelly and form a double-layered triangular lattice.
  The Hamiltonian is given as
        \begin{align}
            \label{eq:spin_hamiltonian}
            \mathcal{H}
        & = \mathcal{H}_{\text{intra}} + \mathcal{H}_{\text{inter}}, \notag \\
        &    \mathcal{H}_{\text{intra}}
         = \sum_{i = 1}^{N}
                \left[
                    J \boldsymbol{S}_{i_{1}} \cdot \boldsymbol{S}_{i_{2}}
                + B \left( \boldsymbol{S}_{i_{1}} \cdot \boldsymbol{S}_{i_{2}} \right)^{2}
                \right], \notag \\
        &    \mathcal{H}_{\text{inter}}
         = \sum_{\braket{i, j}} \sum_{\gamma = 1, 2}
                \left(
                    J' \boldsymbol{S}_{i_{\gamma}} \cdot \boldsymbol{S}_{j_{\gamma}}
                + J''\boldsymbol{S}_{i_{\gamma}} \cdot \boldsymbol{S}_{j_{\bar{\gamma}}}
                \right).
        \end{align}
  where $\boldsymbol{S}_{i_{\gamma}}$ is the spin-1 operator of $\gamma$-th site on a $i$-th dimer. 
  Here, we descriminate $\boldsymbol{S}_{i_{\gamma}}$ defined on a lattice site 
  from $\boldsymbol{\mathcal{S}}_{i}$ which is spin-1 defined on a dimer-bond introduced the next section. 
  The summation $\braket{i, j}$ is taken over all the neighboring pairs of dimers,
  and $\bar{1} = 2$ and $\bar{2} = 1$.
  $J$~$(> 0)$ and $B$~$(> 0)$ denote the antiferromagnetic Heisenberg and the biquadratic interactions,
  respectively, 
  $J'$ and $J''$ are the inter-dimer Heisenberg interactions (Fig.~\ref{fig:lattice}(b)),
  and $N$ is the number of dimers. 
\par
  In \S.~\ref{sec:boson}, we analyze the Hamiltonian Eq.~(\ref{eq:spin_hamiltonian}) by transforming it 
  to the effective model of the $\mathcal{S}=1$ bosons via the perturbation theory, 
  and then solving it by the numerical exact diagonalization (ED) on a finite cluster. 
  The magnetic properties of the model is described by the $\mathcal{S}=1$ carried by the hard-core boson.  
\par
  To disclose the details of the phase diagram, particularly to fix the existence of the long range order, 
  we analyze the structure of the low lying excited states, namely a tower of states in \S.~\ref{sec:lro}. 
  There, the spin-1 bosonic description is found to be not enough to understand several different classes of 
  phases that form a quadrupolar ordering, which are classified as different types of ``nematic'' phases. 
  We therefore introduce different operators defined on a dimer-bond other than $\boldsymbol{\mathcal{S}}_{i}$. 
  To avoid confusion, we separate \S.~\ref{sec:lro} from \S.~\ref{sec:boson} where the latter deals 
  fully with the $\mathcal{S}=1$ bosonic description. 
\par
  The classification of the above mentioned different types of phases in comparison with 
  the previously known phases based on quadrupolar moments is given in \S.~\ref{sec:discussions}, 
  followed by a brief summary in \S.~\ref{sec:summary}. 

\section{Effective \texorpdfstring{$\mathcal{S}=1$}{S=1} bosonic model}
\label{sec:boson}
\subsection{Derivation of the effective Hamiltonian of bosons}
\label{sec:derivation}
\subsubsection{Low-energy states of spin-1 dimers}
\label{subsec:lowenergy_states}
\begin{figure}
			\centering
			\includegraphics[width=86mm]{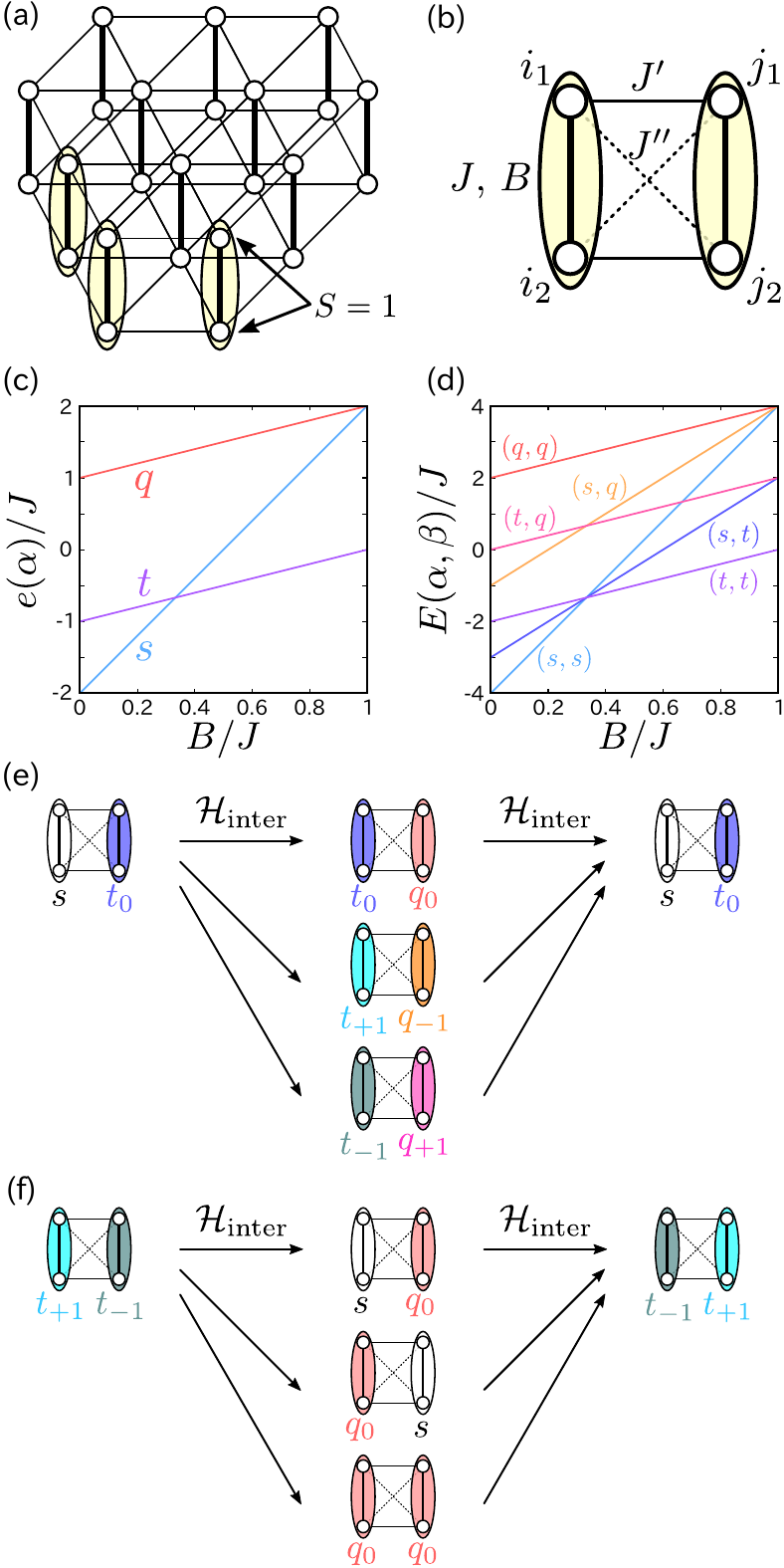}
			\caption{(a) Schematic illustration of the spin-1 dimer model on a triangular lattice.
							(b) Intra- and inter-dimer interactions, where $J>0$, $J'$, $J''$ are the Heisenberg exchanges, and $B$ is the biquadratic exchange.
							(c) Eigenenergy levels of $\mathcal{H}_{\text{intra}}$ of an isolated dimer. 
								$s$, $t$, and $q$ denote the singlet, triplet, and quintet states of the dimer,respectively.
							(d) Eigenenergy levels of $\mathcal{H}_{\text{intra}}$ of two isolated dimers without the inter-dimer $J'$ and $J''$. 
							(e), (f) Examples of the second order perturbation processes. 
							Processes in panel (e) returns to the original state $\ket{s, t_{0}}$ 
							and those in panel (f) exchange the $\mathcal{S}^{z} = +1$ and $\mathcal{S}^{z} = -1$ triplets. 
							They are the origins of the biquadratic interactions between two triplets
							$\left( \boldsymbol{\mathcal{S}}_{i} \cdot \boldsymbol{\mathcal{S}}_{j} \right)^{2}$.
			}
			\label{fig:lattice}
		\end{figure}

  Let us first consider an isolated dimer consisting of two spin-1 interacting via the BLBQ interactions
  $\mathcal{H}_{\text{BLBQ}} = J \boldsymbol{S}_{1} \cdot \boldsymbol{S}_{2}
   + B \left( \boldsymbol{S}_{1} \cdot \boldsymbol{S}_{2} \right)^{2}$.
  The energy eigenstates of $\mathcal{H}_{\text{BLBQ}}$ are classified into singlet~($s$), triplets~($t$), and quintets~($q$),
  and their energies are given as $e(s) = -2J+4B$, $e(t) = -J+B$, and $e(q) = J+B$, respectively.
  Figure~\ref{fig:lattice}(c) shows these energy levels as a function of $B/J$.
  At small $B/J$, the Heisenberg interaction is dominant and the lowest energy state is a singlet.
  This singlet state is replaced by the triplet state when $B/J > 1/3$,
  while the quintet cannot have lower energy than the triplet and remain as the excited states.
\par
  As a starting point of the perturbation, 
  we take $\mathcal{H}_{\text{inter}} = 0$, 
  where the ground state is the product state of the singlets on the isolated dimers for $B/J < 1/3$, 
  and that of triplets for $B/J>1/3$.
  In introducing $\mathcal{H}_{\text{inter}} \ne 0$,
  we consider the processes up to second order in $J'/J$ and $J''/J$, 
  so that the effective interactions between two adjacent dimers appear mainly in the result.
  The energies of the disconnected two dimers with $\alpha$- and $\beta$-multiplets $E(\alpha, \beta)$ are shown in Fig.~\ref{fig:lattice}(d).
  One can see that the states including quintets are higher in energy than the states without quintets when $B/J < 2/3$. 
  Therefore, based on the natural assumption that $B/J$ is small enough, we construct the effective Hamiltonian for the low-energy manifold of states including only singlets and triplets. 
\par
  The first order process contributes to the energy correction of singlet and triplet states, as well as 
  to the exchange of triplet and singlet on the neighboring two dimers. 
  Within the second order perturbation processes between two adjacent dimers,
  the intermediate excited states have at least one quintet
  as shown in the examples of the processes; 
  in Fig.~\ref{fig:lattice}(e), 
  the two-dimer state $\ket{s, t_{0}}$ returns to the same state through the excited states $\ket{t_{0}, q_{0}}$,
  $\ket{t_{+1}, q_{-1}}$ and $\ket{t_{-1}, q_{+1}}$, where $\ket{s}$ is the singlet state,
  and $\ket{t_{\mu}}$ and $\ket{q_{\mu}}$ are the triplet and  the quintet states with $S^{z} = \mu$, respectively.
  In the processes shown in Fig.~\ref{fig:lattice}(f),
  $\ket{t_{+1}, t_{-1}}$, the two-dimer states with $S^{z} = +1$ and $S^{z} = -1$ triplet dimers,
  mixes with $\ket{t_{-1}, t_{+1}}$ via the three excited states 
  $\ket{s_{0}, q_{0}}$, $\ket{q_{0}, s_{0}}$ and $\ket{q_{0}, q_{0}}$.
\par
  The low-energy basis can be described in the spin-1 hard-core bosonic language. 
  The singlet corresponds to the vacuum, 
  and the triplets are the bosons which are not allowed to doubly occupy a dimer.
  This kind of treatment is equivalent to the bond-operator approach,
  developed for the spin-1/2 dimer systems~\cite{Chubukov1989_JETPLett_49_129,Sachdev1990_PhysRevB_41_9323},
  and later applied to spin-1 dimer systems~\cite{Brenig2001_PhysRevB_64_214413,Wang2000_PhysRevB_61_4019} and also to general spin-$S$ dimers~\cite{Kumar2010_PhysRevB_82_054404}.
  We choose the time-reversal invariant form of the basis set $\left\{ \ket{t_{i, \alpha}} \right\}$ described as,
        \begin{align}
            \ket{t_{i, x}} 
        & = \dfrac{\mathrm{i}}{2}
                    \left( \ket{+1, 0} - \ket{0, +1} - \ket{0, -1} + \ket{-1, 0} \right), \notag \\
            \ket{t_{i, y}} 
        & = \dfrac{1}{2}
                    \left( \ket{+1, 0} - \ket{0, +1} + \ket{0, -1} - \ket{-1, 0} \right), \notag \\
            \ket{t_{i, z}} 
        & = - \dfrac{\mathrm{i}}{\sqrt{2}} \left( \ket{+1, -1} - \ket{-1, +1} \right),
        \label{eq:triplets}
        \end{align}
  where the dimer states on the r.h.s. described as $\ket{S_{i_{1}}^{z}, S_{i_{2}}^{z}}$ are those classified by the $S^{z}$-values of the two spins forming a dimer.
  The details of the bond-operator approach 
  and the description of the original spin operators using the bosonic operators are shown in Appendix~\ref{subsec:appendix_bondoperator}.
\subsubsection{Effective Hamiltonian}
\label{subsec:effective_hamiltonian}
  The triplet state with $\alpha$-component $\ket{t_{i, \alpha}}$ at site-$i$ is expressed as $b_{i, \alpha}^{\dag} \ket{0}$,
  where $\ket{0}$ is the singlet state and $b_{i, \alpha}^{\dag}$ is the creation operator of a boson representing that triplet. 
  Using this bosonic operator, the effective Hamiltonian $\mathcal{H}_{\text{eff}}$ up to second order in $J'/J$ and $J''/J$ is given as
        \begin{align}
            \label{eq:boson_hamiltonian}
            \mathcal{H}_{\text{eff}}
        & = E_{0} + \mathcal{H}_{\mu} + \mathcal{H}_{t} + \mathcal{H}_{P} + \mathcal{H}_{V} + \mathcal{H}_{\mathcal{J}} + \mathcal{H}_{\mathcal{B}} + \mathcal{H}_{\text{3body}}, \notag \\
            \mathcal{H}_{\mu}
        & = - \mu \sum_{i = 1}^{N} n_{i} \notag \\
            \mathcal{H}_{t}
        & = t \sum_{\braket{i, j}} \sum_{\alpha = x, y, z}
                    b_{i, \alpha}^{\dag} b_{j, \alpha} + \text{h.c.}, \notag \\
            \mathcal{H}_{P}
        & = P \sum_{\braket{i, j}} \sum_{\alpha = x, y, z}
                    b_{i, \alpha}^{\dag} b_{j, \alpha}^{\dag} + \text{h.c.}, \notag \\
            \mathcal{H}_{V}
        & = V \sum_{\braket{i, j}} n_{i} n_{j}, \notag \\
            \mathcal{H}_{\mathcal{J}}
        & = \mathcal{J} \sum_{\braket{i, j}} 
                \boldsymbol{\mathcal{S}}_{i} \cdot \boldsymbol{\mathcal{S}}_{j} n_{i} n_{j}, \notag \\
            \mathcal{H}_{\mathcal{B}}
        & = \mathcal{B} \sum_{\braket{i, j}} 
                    \left( \boldsymbol{\mathcal{S}}_{i} \cdot \boldsymbol{\mathcal{S}}_{j} \right)^{2} n_{i} n_{j}.
        \end{align}
  Here, $n_{i} = \sum_{\alpha} b_{i, \alpha}^{\dag} b_{i, \alpha}$ is the number operator, 
  and the hard-core condition $n_{i} = 0$ or $1$ is imposed on the number operator. 
  The spin-1 operator of $i$-th boson is expressed by $\boldsymbol{\mathcal{S}}_{i}$,
  where $\mathcal{S}_{i}^{\alpha} = - \mathrm{i} \sum_{\beta, \gamma} \varepsilon_{\alpha\beta\gamma} b_{i, \beta}^{\dag} b_{i, \gamma}$
  and $\varepsilon_{\alpha\beta\gamma}$ is the Levi--Civita symbol.
  We note that the Hamiltonian keeps the $\mathrm{SU}(2)$ symmetry of triplets~\cite{Lecheminant2006_PhysRevB_74_224426,Totsuka2012_PhysRevB_86_014435,Yokoyama2018_PhysRevB_97_180404}, 
  as far as the magnetic field is not applied~\cite{Nikuni2000_PhysRevLett_84_5868}.
\par
  The parameters included in $\mathcal{H}_{\text{eff}}$ is described by the original interaction parameters in Eq.~(\ref{eq:spin_hamiltonian}) as
        \begin{align}
            \label{eq:boson_coefficients}
            &    E_{0} = \left( -2J + 4B \right) N, \notag \\
            &    \mu = - J + 3B
                            + \dfrac{20z}{27 \left( J - B\right)} \left( J' - J'' \right)^{2}, \notag \\
            &    t = \dfrac{4}{3} \left( J' - J''\right), \ 
                P = - \dfrac{4}{3} \left( J' - J''\right), \notag \\
            &    V = \left[ \dfrac{40}{27 \left( J - B \right)} - \dfrac{8}{9 \left( J + 3B \right)}
                                        - \dfrac{2}{9J}
                        \right] \left( J' - J'' \right)^{2}, \notag \\
            &    \mathcal{J} = \dfrac{J' + J''}{2}
                            + \left[ - \dfrac{4}{3 \left( J + 3B \right)} + \dfrac{1}{12J} \right] \left( J' - J'' \right)^{2}, \notag \\
            &    \mathcal{B} = \left[ - \dfrac{4}{9 \left( J + 3B \right)} - \dfrac{1}{144J} \right] \left( J' - J'' \right)^{2},
        \end{align}
  where $z$ is the coordination number. 
  One can immediately see that $\mu, t, P$ and $\mathcal{J}$-terms include the terms that originate from the first order process, 
  whereas $V$ and $\mathcal{B}$-terms do not. 
\par
  There are some processes at the second order level where the three dimers take part in,
  which we denote as $\mathcal{H}_{\text{3body}}$ in Eq.~(\ref{eq:boson_hamiltonian}). 
  We numerically evaluate the effects of $\mathcal{H}_{\text{3body}}$ on the effective Hamiltonian
  by comparing the energies of the ground states of the original Hamiltonian $\mathcal{H}$ (Eq.~(\ref{eq:spin_hamiltonian})),
  and of the effective Hamiltonian $\mathcal{H}_{\text{eff}}$ (Eq.~(\ref{eq:boson_hamiltonian})) 
  with and without $\mathcal{H}_{\text{3body}}$ in a small cluster, 
  finding that it does not play a significant role. 
  We thus discard this $\mathcal{H}_{\text{3body}}$ term in the following for simplicity.
  The details of the evaluation of the effective model is shown in the Appendix~\ref{subsec:appendix_evaluation}. 
  We further show that even the other second order terms included in Eq.~(\ref{eq:boson_hamiltonian}), 
  do not contribute much to the majority of phases we deal with. 
  The way how the inter-dimer interactions work thus turns out to be simple. 
\subsubsection{Physical quantities}
\label{subsec:physical_quantitles}
  For the analysis of the effective model, we calculate the following properties that characterize the ground state. 
  The boson density per dimer is denoted as $\braket{n_{t}}=N^{-1}\sum_{i=1}^{N} \braket{n_{i}}$, 
  and its structure factor is given as 
        \begin{align}
            N(\boldsymbol{k})
        = \dfrac{1}{N} \sum_{i, j = 1}^{N} \braket{n_{i}n_{j}}
                \mathrm{e}^{\mathrm{i}\boldsymbol{k} \cdot \left( \boldsymbol{r}_{i} - \boldsymbol{r}_{j} \right)}. 
        \end{align}
\par
  The magnetic properties are examined by the spin and quadrupole structure factors
        \begin{align}
            \mathcal{S}(\boldsymbol{k})
        &    = \dfrac{1}{N} \sum_{i, j = 1}^{N}
                    \braket{\boldsymbol{\mathcal{S}}_{i} \cdot \boldsymbol{\mathcal{S}}_{j} n_{i}n_{j}}
                    \mathrm{e}^{\mathrm{i}\boldsymbol{k} \cdot \left( \boldsymbol{r}_{i} - \boldsymbol{r}_{j} \right)}, \label{eq:spin_structure_factor} \\
            \mathcal{Q}(\boldsymbol{k})
        &    = \dfrac{1}{N} \sum_{i, j = 1}^{N}
                    \braket{\boldsymbol{\mathcal{Q}}_{i} \cdot \boldsymbol{\mathcal{Q}}_{j} n_{i}n_{j}}
                    \mathrm{e}^{\mathrm{i}\boldsymbol{k} \cdot \left( \boldsymbol{r}_{i} - \boldsymbol{r}_{j} \right)}, \label{eq:quadrupole_structure_factor}
        \end{align}
  where $\boldsymbol{\mathcal{Q}}_{i}$ is the 5-component vector representation of quadrupole operator of spin-1 bosons defined as
        \begin{align}
            \boldsymbol{\mathcal{Q}}_{i}
        = \begin{pmatrix}
                \mathcal{Q}_{i}^{x^{2}-y^{2}} \\ \mathcal{Q}_{i}^{3z^{2}-r^{2}} \\ 
                \mathcal{Q}_{i}^{xy} \\ \mathcal{Q}_{i}^{yz} \\ \mathcal{Q}_{i}^{zx} \\
            \end{pmatrix}
        = \begin{pmatrix}
                \left( \mathcal{S}_{i}^{x} \right)^{2} - \left( \mathcal{S}_{i}^{y} \right)^{2} \\
                \dfrac{1}{\sqrt{3}} 
                    \left[ 3 \left( \mathcal{S}_{i}^{z} \right)^{2} 
                                - \mathcal{S} \left( \mathcal{S} + 1 \right) \right] \\
                \mathcal{S}_{i}^{x} \mathcal{S}_{i}^{y} + \mathcal{S}_{i}^{y} \mathcal{S}_{i}^{x} \\
                \mathcal{S}_{i}^{y} \mathcal{S}_{i}^{z} + \mathcal{S}_{i}^{z} \mathcal{S}_{i}^{y} \\
                \mathcal{S}_{i}^{z} \mathcal{S}_{i}^{x} + \mathcal{S}_{i}^{x} \mathcal{S}_{i}^{z} \\
            \end{pmatrix}.
            \label{eq:quadrupole_operator}
        \end{align}
\par
  In a system with spin-1 defined on each site, typically represented by the spin-1 BLBQ models, 
  the quadrupole operator $\boldsymbol{\mathcal{Q}}_{i}$ is the on-``site'' operator. 
  For a system with spin-1/2 per site, the quadrupole operator is defined on a bond instead, since one needs to prepare a spin-1 from 
  two spin-1/2's~\cite{Shannon2006_PhysRevLett_96_027213,Penc2011_IFM_Chap13_SpinNematics}. 
  In the present case, the two sites forming a dimer each hosts spin-1 operators, $\boldsymbol{S}_{i_{1}}$ and $\boldsymbol{S}_{i_{2}}$, 
  and $\boldsymbol{\mathcal{S}}_{i}$, which is defined on a dimer-bond is a composition of these two spin-1's. 
  For this reason, one can also define another quadrupole operator on dimer-bond as 
        \begin{align}
            Q_{i_{12}}^{\alpha\beta} 
        = S_{i_{1}}^{\alpha} S_{i_{2}}^{\beta} + S_{i_{1}}^{\beta} S_{i_{2}}^{\alpha}
            - \dfrac{2}{3} \left( \boldsymbol{S}_{i_{1}} \cdot \boldsymbol{S}_{i_{2}} \right) \delta_{\alpha\beta}.
        \end{align}
  Then, one finds that $\boldsymbol{Q}_{i_{12}}$ and $\boldsymbol{\mathcal{Q}}_{i}$ are equivalent in terms of our triplet states, namely,
        \begin{align}
            \braket{t_{\alpha}|Q_{i_{12}}^{\mu\nu}|t_{\beta}}
        = \braket{t_{\alpha}|\mathcal{Q}_{i}^{\mu\nu}|t_{\beta}}
        \end{align}
  holds for $\alpha, \beta, \mu, \nu = x, y, z$.
  In the same manner, the spin operator inside $i$-th spin-1 dimer defined as
        \begin{align}
            S_{i_{12}}^{\alpha} = S_{i_{1}}^{\alpha} + S_{i_{2}}^{\alpha}
        \end{align}
  works in the same way as $\mathcal{S}_{i}^{\alpha}$ for the triplet states, i.e.,
        \begin{align}
            \braket{t_{\alpha}|S_{i_{12}}^{\mu}|t_{\beta}} 
        = \braket{t_{\alpha}|\mathcal{S}_{i}^{\mu}|t_{\beta}}
        \end{align}
  holds for $\alpha, \beta, \mu = x, y, z$.
\par
  Unlike the spin-1 BLBQ models~\cite{Lauchli2006_PhysRevLett_97_087205}, 
  the number of spin-1 bosons per dimer is not fixed in our Hamiltonian. 
  However, one can consider the spin-1 BLBQ model as the $\braket{n_{t}}=1$-limiting case of our model since the two models share the same definition, Eq.~(\ref{eq:quadrupole_operator}). 
  One can thus make use of the analysis applied to the spin-1 BLBQ model~\cite{Lauchli2006_PhysRevLett_97_087205}; 
  there are so-called $\mathrm{SU}(3)$-points in the BLBQ model, 
  where the three components of $\boldsymbol{\mathcal{S}}$ and the five components of $\boldsymbol{\mathcal{Q}}$ 
  equivalently form the eight elements of the $\mathrm{SU}(3)$ Lie algebra. 
  Exactly at this point the transition between the magnetic and the spin nematic phases is known to take place. 
  Numerically, this transition is identified by the point where $\mathcal{S}(\boldsymbol{k})$ and $\bar{\mathcal{Q}}(\boldsymbol{k})\equiv (3/5)\mathcal{Q}(\boldsymbol{k})$ 
  take the same values. We thus use this normalized value $\bar{\mathcal{Q}}(\boldsymbol{k})$ to determine the phase transitions between the magnetic and the quadrupolar states. 
    \begin{figure}
      \centering
      \includegraphics[width=86mm]{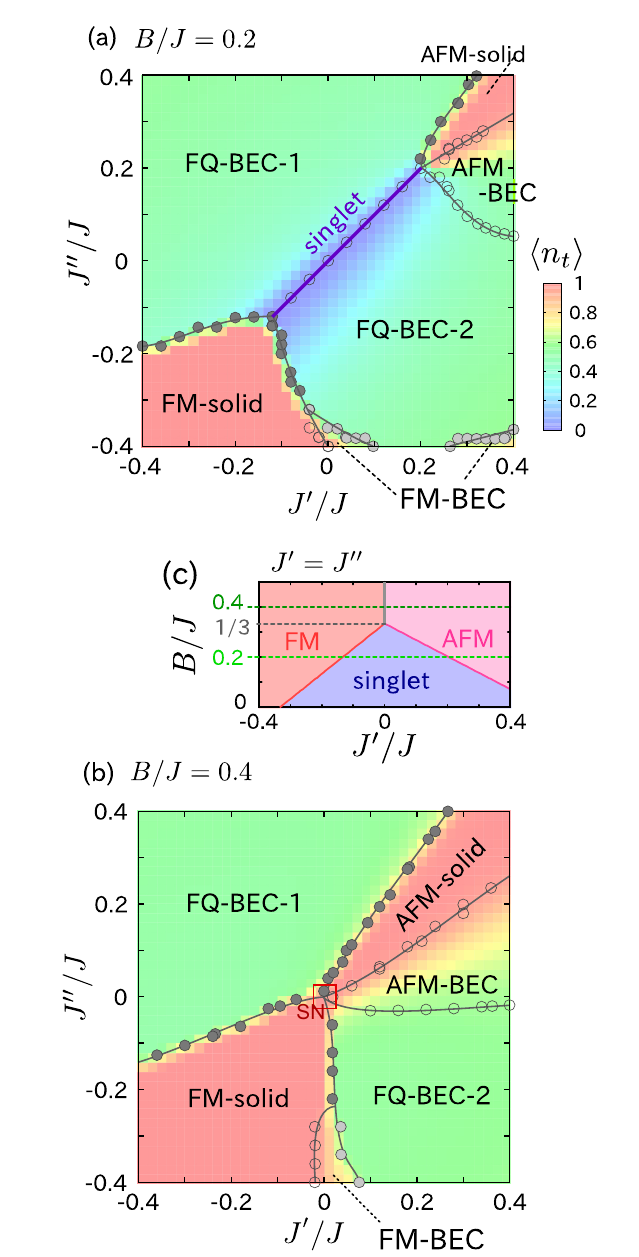}
      \caption{
                Ground state phase diagram of the spin-1 dimer triangular lattice at (a) $B/J = 0.2$ and (b) $0.4$,
                obtained by the numerical diagonalization of the effective model with $N = 12$. 
		            Filled and open circles represent the first and second order phase transitions, 
		            where the transition between FM-BEC and FQ-BEC-2 is weakly first order. 
                FM, AFM, FQ represent the ferromagnetic, antiferromagnetic and ferro-quadrupolar phases, 
                and $\braket{n_{t}}\approx 1$ and $\lesssim 0.9$ are the solid and BEC states of bosons. 
                Colors in the phase diagram are the density plot of the triplet number $\braket{n_{t}}$. 
                The small $J'$, $J''$ region marked with red square in (b) encloses SN phase (see Fig.~\ref{fig:smallphased}(b)).
                (c) Phase diagram on the plane of $J'/J$ and $B/J$, whose fixed $B/J$-lines correspond to 
                the $J'=J''$ line of the phase diagrams in panels (a) and (b). 
      }
      \label{fig:phasediagram_2ndorder}
    \end{figure}
    \begin{figure}
      \centering
      \includegraphics[width=86mm]{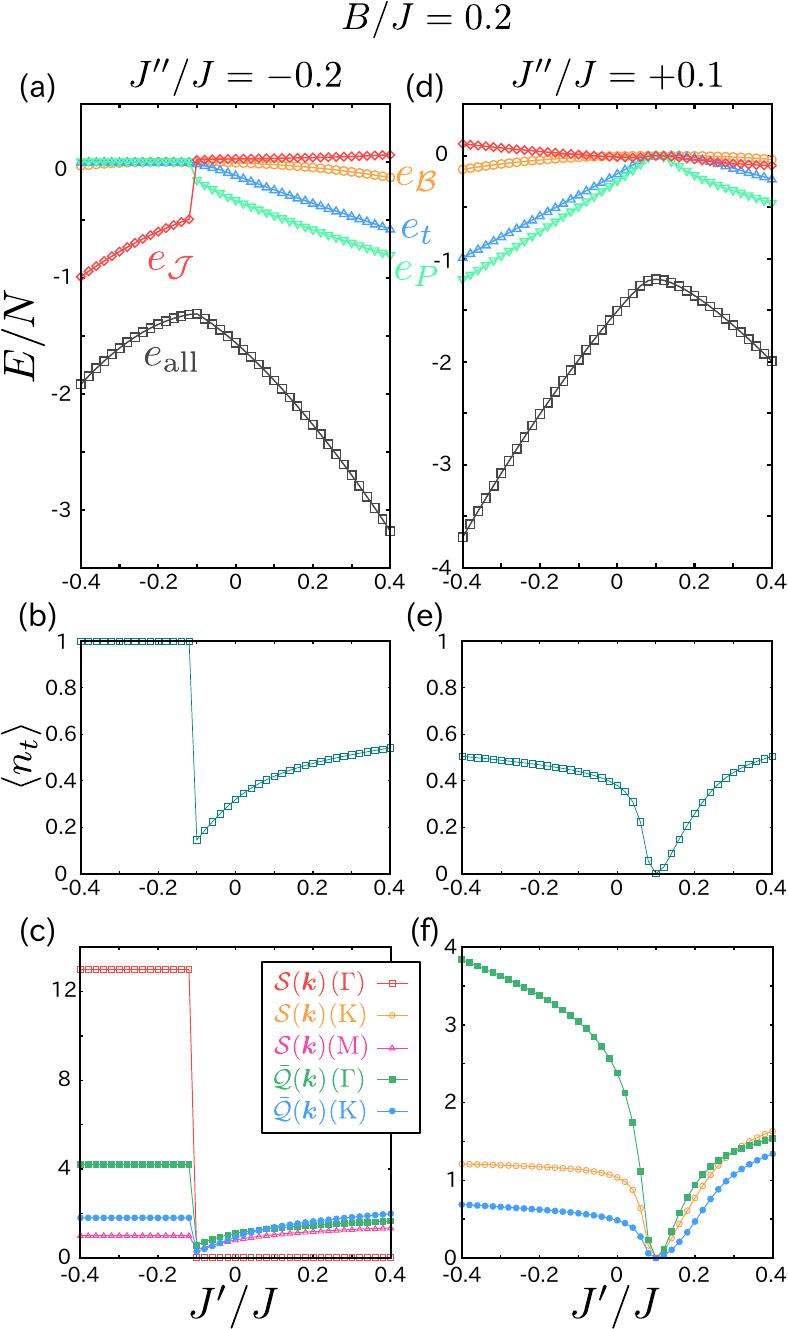}
      \caption{
                $J'/J$ dependences of the physical quantities at $B/J = 0.2$
                when (a)--(c) $J''/J = -0.2$ and (d)--(f) $J''/J = +0.1$.
                (a), (d) Total energies $e_{\text{all}}$ and the contributions from 
                major terms in the effective Hamiltonian, $e_{t}, e_{P}, e_{\mathcal{J}}$ and $e_{\mathcal{B}}$. 
                (b), (e) Triplet densities $\braket{n_{t}}$. 
                (c), (f) Spin (Eq.~(\ref{eq:spin_structure_factor})) and quadrupole (Eq.~(\ref{eq:quadrupole_structure_factor})) 
                structure factors at $\Gamma$, $\mathrm{K}$, $\mathrm{M}$-points in the reciprocal space.
                Quadrupole structure factors denoted as $\bar{\mathcal{Q}}(\boldsymbol{k})$ are normalized 
                to be compared with the spin structure factors $\mathcal{S}(\boldsymbol{k})$.
      }
      \label{fig:physicalquantities_bb020}
   \end{figure}
\subsection{Results of the \texorpdfstring{$\mathcal{S}=1$}{S=1} bosonic model}
\label{sec:results}
\subsubsection{Phase diagram}
\label{subsec:phase_diagram}
  We numerically diagonalize $\mathcal{H}_{\text{eff}}$ on the $N = 12$ triangular lattice ($z = 6$) under the periodic boundary condition. 
  The phase diagrams on the plane of $J'/J$ and $J''/J$ at $B/J = 0.2$ and $0.4$ are shown 
  in Figs.~\ref{fig:phasediagram_2ndorder}(a) and ~\ref{fig:phasediagram_2ndorder}(b).
\par    
  The phase diagram is divided into four parts in overall. 
  When $J'\sim J'' >0$, the antiferromagnetic phases with $\braket{n_{t}}\approx 1$ (AFM-solid) and $\braket{n_{t}}\lesssim 0.9$ (AFM-BEC) are stabilized by the antiferromagnetic interaction, 
  $\mathcal{J}>0$, between bosons occupying the neighboring dimers. 
  On the opposite part of the phase diagram, $J'\sim J' <0$, the ferromagnetic phase with $\braket{n_{t}}\approx 1$ 
  (FM-solid) is realized for the similar reasons. 
  When $J'-J'' < 0$ and $J'-J'' > 0$, 
  two different types of ferroquadrupolar (FQ) phases, FQ-BEC-1 and FQ-BEC-2 
  appear over a wide parameter region. 
  Throughout both of the phase diagrams, we see no particular features of bosons, 
  i.e., $N(\boldsymbol{k})$ takes the maximum value at $\Gamma$-point, 
  which indicates that bosons distribute uniformly in space and 
  does not show any translational symmetry breaking long range order. 
\subsubsection{\texorpdfstring{$J' = J''$}{J' = J''} line}
\label{subsec:diagonal}
  The starting point is $J'=J''=0$, at which the ground state is the product state of the isolated-dimer state. 
  As one can see from Eq.~(\ref{eq:boson_coefficients}), most of the parameters, namely $t,P,V,\mathcal{B}$ 
  are the linear or the square functions of $(J'-J'')$. 
  Therefore, these parameters remain zero exactly at $J'=J''$, namely the inter-dimer interactions cancel out 
  because of the geometrical frustration effect. 
\par
  In fact, when $B/J = 0.2$, the singlet product state, namely $\braket{n_{t}}=0$, 
  remains a ground state along this line. 
  The endpoint of this singlet phase is evaluated in the following manner; 
  when $J' = J''$, the effective Hamiltonian consists only of two terms 
        \begin{align}
            \mathcal{H}_{J'=J''}
        = -\mu \sum_{i = 1}^{N} n_{i} 
        + \mathcal{J}\sum_{\braket{i, j}} \boldsymbol{\mathcal{S}}_{i} \cdot \boldsymbol{\mathcal{S}}_{j} n_{i} n_{j}, 
        \end{align}
  with $\mu = -J + 3B$ and $\mathcal{J} = J'$. 
  Regardless of its sign, $\mathcal{J}$ works as an effective attractive interaction between bosons, 
  since it is energetically favorable to occupy the neighboring pairs of dimers with triplets to gain the magnetic interaction energy. 
  Then, there is a first order transition between the $\braket{n_{t}}=0$-singlet and the $\braket{n_{t}}=1$-FM or AFM solid phases. 
  The phase boundary can be obtained by comparing their energies, $E_{0}(N)$ and $E_{1}(N)$, where there is a relationship, 
        \begin{align}
            E_{1}(N) = E_{0}(N) -\mu N + 3N e_{\text{bond}}. 
        \end{align}
  Here, $e_{\text{bond}}$ is evaluated as the bond-energy of the ground state of the spin-1 triangular lattice Heisenberg model 
  $J'\sum_{\braket{i, j}}\boldsymbol{S}_{i} \cdot\boldsymbol{S}_{j}$ for $N=12$. 
  Figure~\ref{fig:phasediagram_2ndorder}(c) shows the resultant phase diagram on the plane of $J'=J''$ and $B$ with $J=1$. 
  The singlet phase corresponding to the straight line in Fig.~\ref{fig:phasediagram_2ndorder}(a) 
  shrinks toward smaller $J'=J''$ value with increasing $B/J$, and disappears at $B/J=1/3$. 
  For $B/J>1/3$, $\braket{n_{t}}=1$ is realized throughout the whole $J'=J''$ line. 
\subsubsection{Ferromagnetic and antiferromagnetic phases}
\label{subsec:magnetic_phases}
  The FM and AFM phases extend from the endpoints of the $J' = J''$-singlet phase discussed above. 
  In Fig.~\ref{fig:physicalquantities_bb020}, 
  we show the total energy $e_{\text{all}}$ and the contributions from each terms, $e_{t}$, $e_{P}$, $e_{\mathcal{J}}$, and $e_{\mathcal{B}}$, 
  the boson density, $\braket{n_{t}}$, and the values of the structure factors at $\Gamma$, $\mathrm{K}$ and $\mathrm{M}$ points of the Brillouin zone. 
  We vary $J'/J$ along the fixed $J''/J=-0.2$ and $0.1$ lines. 
  In the former case, a jump in the physical quantity is found at the transition from the FM-solid to the FQ-BEC-2 phase. 
  Compared to other phases, the FM-solid phase has a large energy gain of $e_{\mathcal{J}}$, indicating that the magnetic interaction $\mathcal{J}$ is responsible for stabilizing the FM-solid.
  Indeed, $\mathcal{S}(\boldsymbol{k})$ shows a peak at the $\Gamma$-point in this phase 
  while the other $\mathcal{S}(\boldsymbol{k})$ and $\mathcal{Q}(\boldsymbol{k})$ remain small. 
  When we vary $J'/J$ along $J''/J=0.1$ (Figs.~\ref{fig:physicalquantities_bb020}(d)--(f)), 
  $\braket{n_{t}}\lesssim 0.55$, and the system remains a BEC. 
  The transitions along this line are of second order. 
  At $J'/J\gtrsim 0.3$, $\mathcal{S}(\boldsymbol{k})$ at $\mathrm{K}$-point starts to overwhelm $\mathcal{Q}(\boldsymbol{k})$ at the $\Gamma$-point 
  which we recognize as the AFM-BEC phase, following the treatment in Ref.[~\onlinecite{Lauchli2006_PhysRevLett_97_087205}] (see the last part of \S.~\ref{subsec:physical_quantitles}). 
  The phase boundaries in Fig.~\ref{fig:phasediagram_2ndorder} are classified into first and second order ones (filled and open circles) 
  according to this analysis. 
\subsubsection{FQ phases}
\label{subsec:spinnematic_phases}
  In the phase diagram, there are two different ferroquadrupolar phases, FQ-BEC-1 extending at $J''>J'$ and FQ-BEC-2 at $J''<J'$. 
  As we see in Figs.~\ref{fig:physicalquantities_bb020}(e) and \ref{fig:physicalquantities_bb020}(f), 
  $\braket{n_{t}}$ and $\mathcal{Q}(\boldsymbol{k})$ both once decrease down to zero at the boundary of the two phases 
  where the singlet state appears, which marks the second order transition. 
  In such a case, the order parameters of the two phases should differ. 
  In fact, although $\mathcal{Q}(\boldsymbol{k})$ at $\Gamma$-point is dominant in both phases, only in the FQ-BEC-2 phase 
  $\mathcal{Q}(\boldsymbol{k})$ at $\mathrm{K}$-point takes as large value as well.
\par
\begin{figure}
      \centering
      \includegraphics[width=86mm]{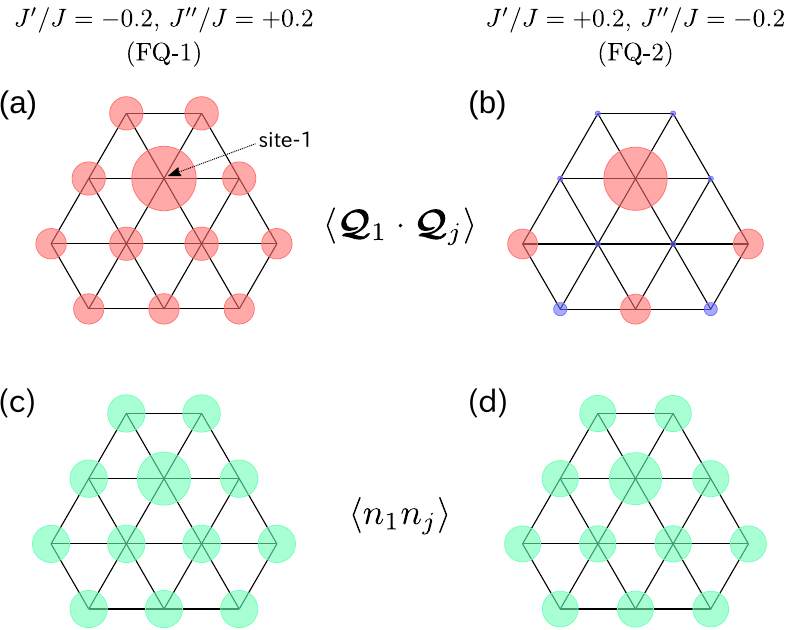}
      \caption{Spatial correlation functions for $(J'/J, J''/J)= (-0.2, +0.2)$ in FQ-1 
               and $(+0.2, -0.2)$ in FQ-2 phases at $B/J = 0.2$; 
               the quadrupolar correlations in (a), (b) $\braket{\boldsymbol{\mathcal{Q}}_{1} \cdot \boldsymbol{\mathcal{Q}}_{j}}$ 
               and the boson-boson correlation in (c), (d) $\braket{n_{1}n_{j}}$. 
               Areas of the circles are proportional to the amplitude of correlations, 
               $\left| \braket{\boldsymbol{\mathcal{Q}}_{1} \cdot \boldsymbol{\mathcal{Q}}_{j}} \right|$ 
               or $\left|\braket{n_{1}n_{j}}\right|$.
               Red and blue circles in (a), (b) correspond to the signs of $\braket{\boldsymbol{\mathcal{Q}}_{1} \cdot \boldsymbol{\mathcal{Q}}_{j}}$, positive and negative, respectively.
      }
      \label{fig:correlation_circle}
    \end{figure}

  Figures~\ref{fig:correlation_circle}(a) and \ref{fig:correlation_circle}(b) show 
  the two-point quadrupole correlations between site-1 and site-$j$, 
  $\braket{\boldsymbol{\mathcal{Q}}_{1}\cdot\boldsymbol{\mathcal{Q}}_{j}}$, 
  in FQ-BEC-1 and FQ-BEC-2 phases. 
  The former correlation develops uniformly in space, whereas in the latter, 
  there is apparently a growth of correlation in the period of twice the lattice spacing in all three directions. 
  This three-sublattice-like structure of quadrupole moments corresponds to the peak of $\mathcal{Q}(\boldsymbol{k})$ at the $\mathrm{K}$-point. 
  Figures~\ref{fig:correlation_circle}(c) and \ref{fig:correlation_circle}(d) 
  are the two-point correlation of bosons, $\braket{n_{1}n_{j}}$, which are both uniform in space. 
  This indicates that the three-sublattice structure of the quadrupolar moment in the FQ-BEC-2 is not because of 
  the modulated the bosonic distribution but originates purely from the correlation between the spin degrees of freedom $\boldsymbol{S}_{i}$; 
  the nearest-neighbor quadrupolar correlation is suppressed, 
  while the next nearest-neighboring correlations are ferroic. 
\subsubsection{Case of \texorpdfstring{$B/J = 0.4$}{B/J = 0.4}}
\label{subsec:results_bb040}
\begin{figure}
 	\centering
 	\includegraphics[width=86mm]{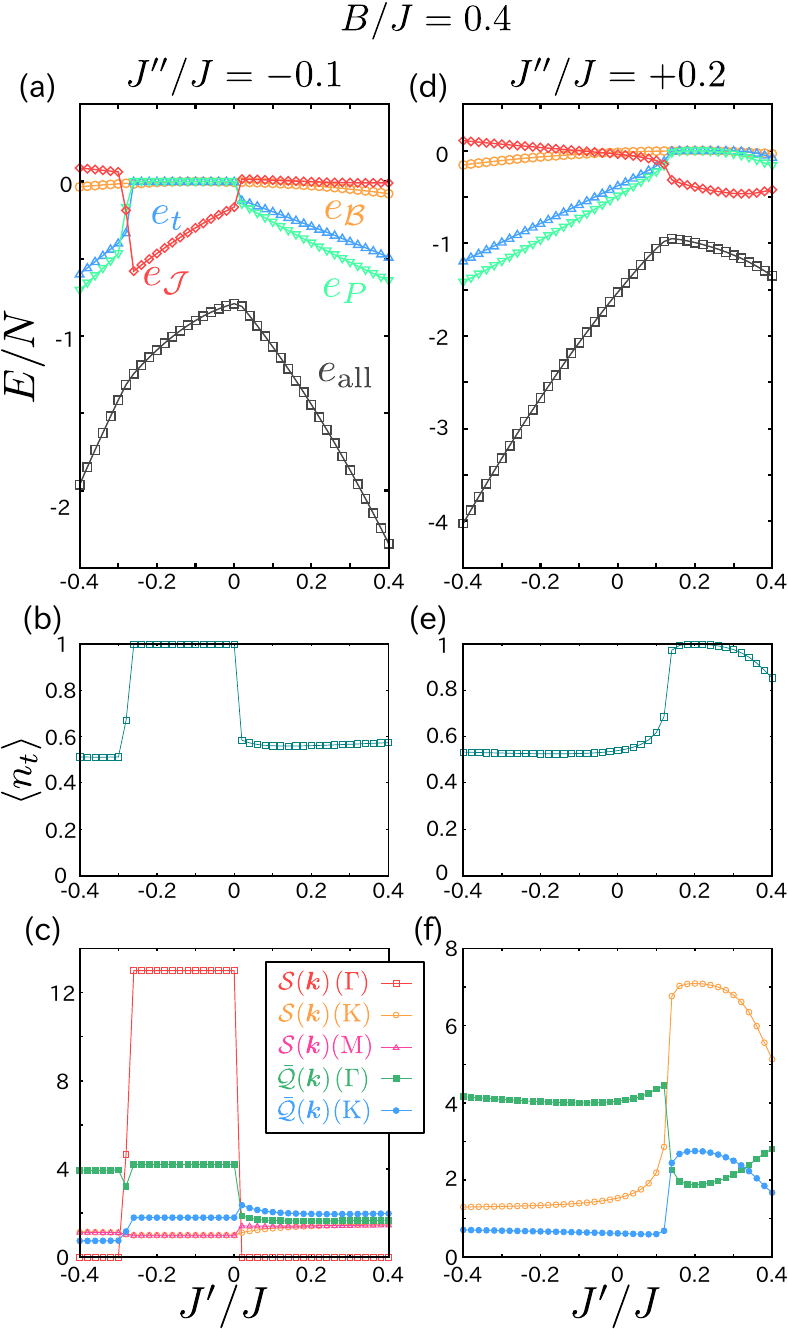}
 	\caption{
 		$J'/J$ dependence of the physical quantities at $B/J = 0.4$
 		when (a)--(c) $J''/J = -0.1$ and (d)--(f) $J''/J = +0.2$.
 		(a), (d) Total energies and partial energies of some terms of the effective Hamiltonian.
 		(b), (e) Triplet densities.
 		(c), (f) Spin (Eq.~(\ref{eq:spin_structure_factor})) and quadrupole (Eq.~(\ref{eq:quadrupole_structure_factor})) structure factors of some points in the reciprocal space.
 		The qudrupole structure factors are normalized to $\bar{\mathcal{Q}}(\boldsymbol{k})$ 
 		to be compared with the spin structure factors $\mathcal{S}(\boldsymbol{k})$. 
 	}
 	\label{fig:physicalquantities_bb040}
 \end{figure}

  We now focus on the case of $B/J = 0.4$, where $\mu$ takes a positive value. 
  The singlet phase no longer exists and the triplet product state realized at $J'=J''=0$ immediately transforms to either 
  of the phases we discussed earlier when the inter-dimer interactions become finite. 
  Figure~\ref{fig:physicalquantities_bb040} shows the $J'/J$ dependences of energies, 
  boson density, and the structure factors to be compared with Fig.~\ref{fig:physicalquantities_bb020}. 
  The first order transitions separating the FM-solid from FQ phases are observed. 
  The boson density in the FQ-BEC phase remains quite stable at around $\braket{n_{t}}\approx 0.55$, 
  indicating that the nature of the BEC phases does not change much with $B/J$.
\subsubsection{The orders of perturbation}
\label{subsec:perturbation_order}
\begin{figure}
      \centering
      \includegraphics[width=86mm]{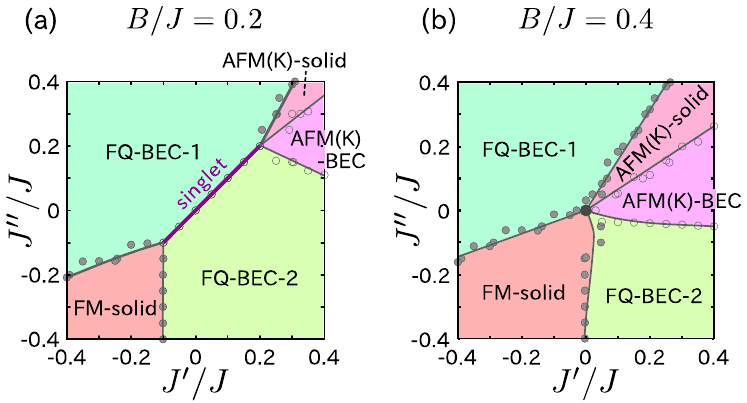}
      \caption{
                (a), (b) Phase diagrams of the effective model up to the first order in $J'/J$ and $J''/J$
                at (a) $B/J = 0.2$ and (b) $B/J = 0.4$,
                obtained by the analysis of the results of the numerical diagonalization on the $N = 12$ cluster. 
            }
      \label{fig:phasediagram_1storder}
    \end{figure}

  The interaction parameters in Eq.~(\ref{eq:boson_coefficients}) include the first and second order terms. 
  Among them, $V$ and $\mathcal{B}$ disappear when we neglect the second order terms. 
  To see how much the second order terms contribute to the determination of the phase diagram, 
  we perform the numerical diagonalization by limiting the parameter values to those up to the first order in $J'/J$ and $J''/J$ with $N=12$. 
  Figures~\ref{fig:phasediagram_1storder}(a) and ~\ref{fig:phasediagram_1storder}(b) show the phase diagrams at $B/J = 0.2$ and $0.4$. 
  The phase diagrams are in good agreement with those in Figs.~\ref{fig:phasediagram_2ndorder}(a) and (b), 
  indicating that $V$ and $\mathcal{B}$ do not play a major role in the five representative phases, 
  FM-solid, AFM-solid/BEC, FQ-BEC-1 and FQ-BEC-2. 
\section{Long range orders}
\label{sec:lro}
\begin{figure*}
      \centering
       \includegraphics[width=172mm]{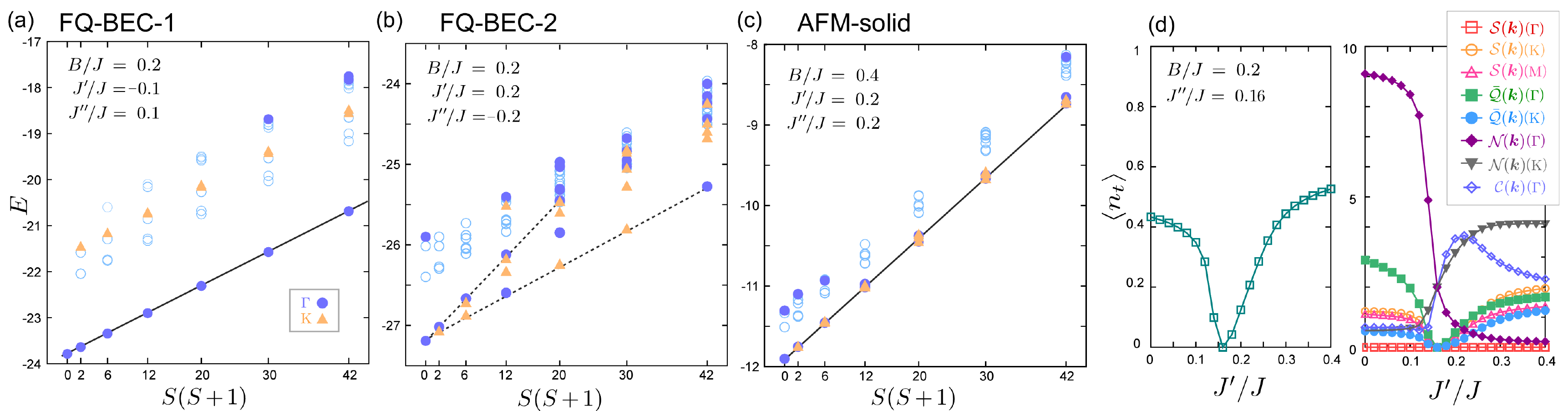}
      \caption{Low-energy excited states as a function of $S(S+1)$ 
                for (a) FQ-BEC-1, (b) FQ-BEC-2, (c) AFM-solid phases for $N = 12$ cluster. 
                Filled circles and triangles are the $\Gamma$ and $K$ points of the Brillouin zone 
                and open circles are from the other $\boldsymbol{k}$-points. 
                Those following the solid line are tower of states indicating the long range order. 
                (d) Structure factors at $\Gamma$, $\mathrm{K}$-points,
                including $\mathcal{N}(\boldsymbol{k})$ and $\mathcal{C}(\boldsymbol{k})$ which gives the information on the 
                details of the magnetic properties of the dimerized spin-$\boldsymbol{S}_{i_{\gamma}}$. 
                 $J'/J$ is varied from FQ-BEC-1 ($J'/J<0.15$), 
                 FQ-BEC-2 ($0.15<J'/J< 0.25$) to AFM-BEC phases. 
            }
      \label{fig:tos}
    \end{figure*}

We so far disclosed the overall magnetic properties of the system in the $\mathcal{S}=1$ bosonic description. 
Since the phase diagrams is obtained by correlation functions in a small finite cluster, 
the nature of the collective ground state, particularly of the FQ phases, should be explicitly determined. 
For that purpose, we perform the thick-restart Lanczos method~\cite{Wu2000_SIAMJMatrixAnalAppl_22_602} 
and disclose the scaling properties of the low-lying levels of the exact spectra. 
Figures~\ref{fig:tos}(a)--\ref{fig:tos}(c) show the ones from the FQ-BEC-1, FQ-BEC-2 and AFM-solid phases, respectively. 
For the (a) FQ-BEC-1 and (c) AFM-solid phases, we clearly find quasidegenerate joint states (QDJS)~\cite{Anderson1952_PhysRev_86_694,Bernu1992_PhysRevLett_69_2590,Bernu1994_PhysRevB_50_10048} that indicate the existence of SU(2) symmetry broken 
long range order marked with solid lines. 

\begin{figure}
      \centering
      \includegraphics[width=86mm]{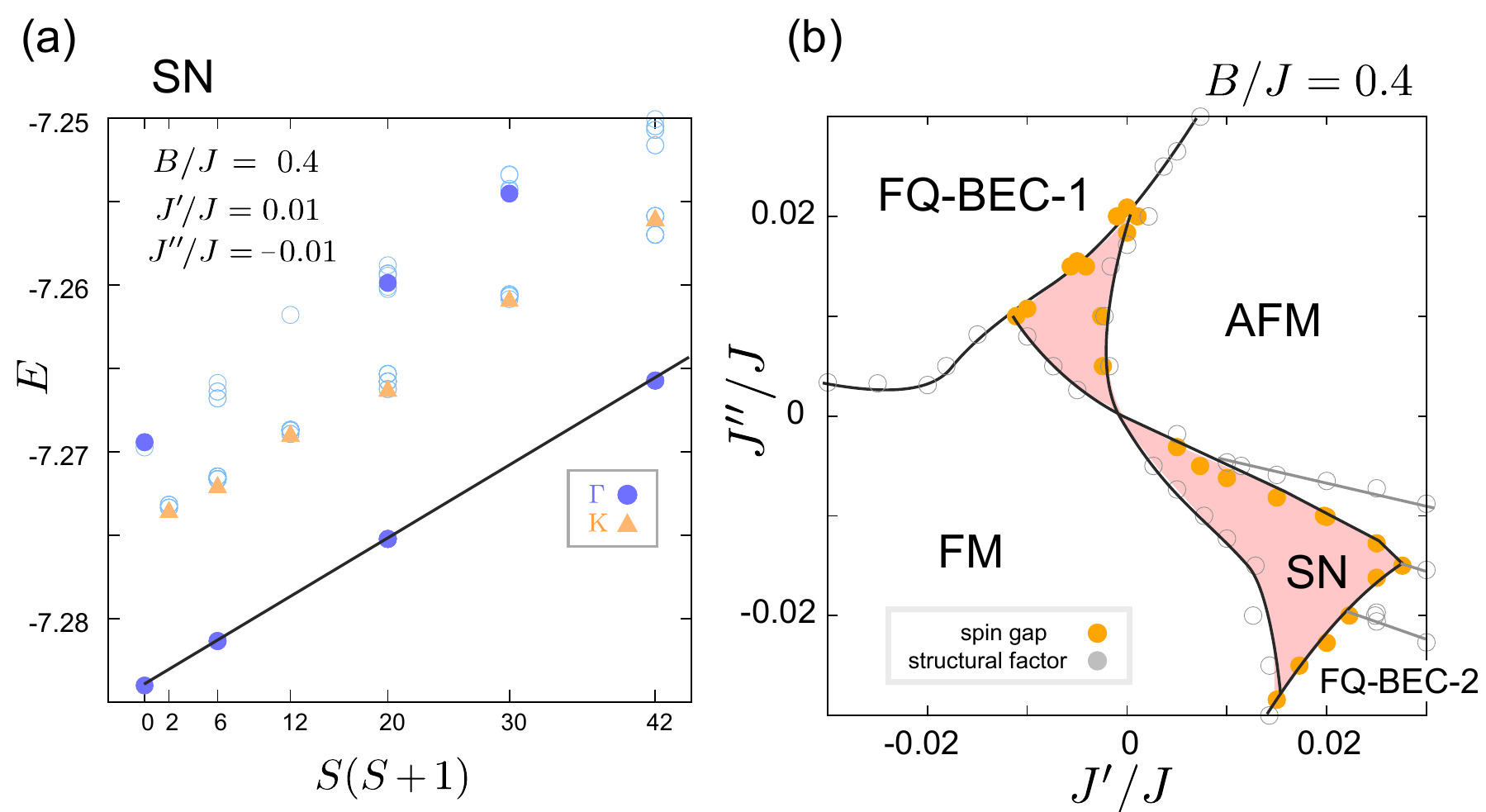}
      \caption{ (a) Low-energy excited states as a function of $S(S+1)$ at $J'/J=0.01$ and $J''/J=-0.01$, $B/J = 0.4$. 
                (b) Phase diagram at small $J'/J$ and $J''/J$ and $B/J = 0.4$ determined by the Anderson tower analysis, 
                which is the magnification of the region indicated by red square 
                at the center part of Fig.~\ref{fig:phasediagram_2ndorder}(c). 
                SN phase is found for $J'\sim -J'$, where the lowest excited state is $S=2$. 
                The boundaries are given both from the crossing of $S=1,2$ levels (filled circle) 
                and from the structural factor (open circle). 
		            Physical quantities along the fixed $J''/J=0.01$ line in the phase diagram
                is given in Fig.~\ref{fig:app_sn}. 
            }
      \label{fig:smallphased}
    \end{figure}

The QDJS found in Fig.~\ref{fig:tos} (a) consists of $S=0,1,2,3,\cdots$ states 
with only one level in each $S$-sector all belonging to $\Gamma$-point, 
characterizing the excitation of the U(1) uniaxial rotator. 
They clearly differ from the one known for the spin-1 nematic phase~\cite{Penc2011_IFM_Chap13_SpinNematics} 
(cf. see Fig.~\ref{fig:smallphased}(a)). 
This FQ-BEC-1 phase is identified as the one found in the $S=1/2$ spin ladder system~\cite{Totsuka2012_PhysRevB_86_014435} 
denoted as F-nematic phase. 
In this phase, if we shift from the bosonic picture to the original pair of spin-$\boldsymbol{S}_{i_{\gamma}}$ forming a dimer, 
one can redescribe state in terms of different operators defined on a dimer; 
the staggered operator of the two spin and the vector chiral operator 
\begin{align}
\label{eq:q}
	\boldsymbol{q}_{i} 
	& = \dfrac{1}{2}\left( \boldsymbol{S}_{i_{1}}- \boldsymbol{S}_{i_{2}} \right), \\
\label{eq:p}
	\boldsymbol{p}_{i} 
	& = \boldsymbol{S}_{i_{1}} \times \boldsymbol{S}_{i_{2}}
\end{align}
following Eq.~(15) in Ref.~\onlinecite{Totsuka2012_PhysRevB_86_014435}, 
which is related to our $\mathcal{S}=1$ bosonic operator as 
$b_{i,\alpha}\propto q_{i}^{\alpha} - \mathrm{i} p_{i}^{\alpha}$ and $b_{i,\alpha}^{\dagger}\propto q_{i}^{\alpha} + \mathrm{i}p_{i}^{\alpha}$. 
The structural factor of $\boldsymbol{q}_{i}$ and $\boldsymbol{p}_{i}$ are 
\begin{align}
\label{eq:staggered_spin_structurefactor}
	\mathcal{N}(\boldsymbol{k})
& = \dfrac{1}{N} \sum_{i,j=1}^{N} 
		\braket{\boldsymbol{q}_{i}\cdot\boldsymbol{q}_{j}}
		\mathrm{e}^{\mathrm{i}\boldsymbol{k}(\boldsymbol{r}_{i}- \boldsymbol{r}_{j})}, \\
	\mathcal{C}(\boldsymbol{k})
& = \dfrac{1}{N} \sum_{i,j=1}^{N}
		\braket{\boldsymbol{p}_{i}\cdot\boldsymbol{p}_{j}}
		\mathrm{e}^{\mathrm{i}\boldsymbol{k}(\boldsymbol{r}_{i}-\boldsymbol{r}_{j})}. 
\end{align}
Since the quadrupolar moment on a dimer-bond is rewritten in the form
\begin{align}
  \mathcal{Q}_{i}^{\alpha\beta}
 = - \left\{
       \frac{3}{4} 
         \left[ \left( q_{i}^{\alpha} q_{i}^{\beta} + p_{i}^{\alpha} p_{i}^{\beta} \right)
              + \left( q_{i}^{\beta} q_{i}^{\alpha} + p_{i}^{\beta} p_{i}^{\alpha} \right) \right]
     - \frac{2}{3} \delta_{\alpha\beta}
     \right\} n_{i},
\end{align}
$\braket{\mathcal{Q}_{i}} \ne 0$ indicates that at least 
$\braket{q_{i}} \ne 0$ or $\braket{p_{i}} \ne 0$ is fulfilled. 
In this description, the F-nematic phase is the ferroic order of staggered operator $\braket{\boldsymbol{q}_{i}\cdot\boldsymbol{q}_{j}} > 0$ and $\braket{p_{i}}=0$, breaking SU(2) down to U(1). 
This means that the small moments induced on these two spins are always forming an antiparallel state 
and \textit{keep the dimer unit always nonmagnetic}, 
whereas these small moments form a ferromagnetic long range order inside the upper and lower 2D layers. 
We confirmed by comparing with the $N=9$ QDJS that the tower of states collapses to the ground state as $1/N$. 
\par
Figure~\ref{fig:tos}(d) shows the boson density $\braket{n_{t}}$ and structural factors 
when $J'/J$ is varied from FQ-BEC-1 ($J'/J<0.15$), FQ-BEC-2 ($0.15<J'/J< 0.25$) to AFM-BEC phases. 
Large $\mathcal{N}(\boldsymbol{k})$ at the $\Gamma$-point characterizes FQ-BEC-1, 
$\mathcal{C}(\boldsymbol{k})$ as well as $\mathcal{N}(\boldsymbol{k})$ at the $\mathrm{K}$-point take large values in FQ-BEC-2 
and particularly $\mathcal{C}(\boldsymbol{k})$ differenciates FQ-BEC-2 from AFM phase. 
\par
We now go back to Fig.~\ref{fig:tos}(b) for FQ-BEC-2 with a QDJS-like structure. 
The states between two broken lines may form a set of ground state manifold with $(2S+1)$-degeneracy 
for each $S$-sector, in consistency with the biaxial rotator or a quantum top 
as found in the 120$^\circ$-N\'{e}el ordering in the spin-1/2 triangular lattice antiferromagnet~\cite{Bernu1992_PhysRevLett_69_2590,Bernu1994_PhysRevB_50_10048}. 
However, these low lying levels are not well separated from the states above them. 
It can be contrasted from the AFM-solid phase with distinct $(2S+1)$-degenerate levels forming QDJS 
which is a clear indication of 120$^\circ$-N\'{e}el ordering of ${\mathcal S}=1$ moments. 
In FQ-BEC-2, one can also focus on the lowests level in each sector near the lower broken lines; 
their $S=0,3,6$ levels at $\Gamma$-point and $S=1,2,4,5$ at $\mathrm{K}$-point are 
symmetric under the symmetry action when putting a uniaxial rotator on a triangular three sublattices, 
suggesting the $p$-nematic type of property~\cite{Lauchli2005_PhysRevLett_95_137206}. 

Again refering to the phase diagram of the spin ladder system in Ref.~\onlinecite{Totsuka2012_PhysRevB_86_014435}, 
there is a similar phase for $J'>0$ called ``NAF (N\'{e}el AF)-dominant'' characterized by antiferroic $\braket{\boldsymbol{q}_{i}\cdot\boldsymbol{q}_{j}} < 0$ and $\braket{p_{i}}=0$.  
Similarly, one may expect some sort of 120$^\circ$-long range order 
of $\braket{q_{i}}$ compatible with the triangular lattice geometry, 
which may be a combined 120$^\circ$-order of $\boldsymbol{S}_{i_{\gamma}}$ in each layer, 
while keeping the dimer nonmagnetic. 
However, in FQ-BEC-2 both $\boldsymbol{q}_{i}$ and $\boldsymbol{p}_{i}$ show large correlation (Fig.~\ref{fig:tos}(d), 
which is not a simple 120$^\circ$-$\braket{q_{i}}$-ordering. 
These results are in agreement with the QDJS, that encloses both the 120$^\circ$-Ne\'el and the $p$-nematic properties. 
One possibility is that the in-plane 120$^\circ$-N\'{e}el ordering of small $\boldsymbol{S}_{i_{\gamma}}$ exists 
but is not fully face to face between layers. 
Whereas, considering the fact that 120$^\circ$-N\'{e}el ordering is relatively subtle due to small moment 
even a standard spin-1/2 triangular lattice Heisenberg model~\cite{Mezzacapo2010_NewJPhys_12_103039}, 
such kind of ordering may not be stable as a long range order in our system 
where the activated local magnetic moment is expected to be small, and may even be masked by the 
intra-dimer quantum fluctuation, in which case the pure $p$-nematic ordering may be stabilized. 
\par
Besides these two FQ-phases we find a conventional spin nematic (SN) phase at $B/J \gtrsim 1/3$ 
in a small region of $J'/J$ and $J''/J$, which can only be 
detected by the Anderson tower analysis. 
Figure~\ref{fig:smallphased}(a) shows the QDJS consisting of $S=0,2,4,\cdots$, 
which indicates the SU(2)-symmetry-broken SN on a triangular lattice~\cite{Penc2011_IFM_Chap13_SpinNematics}. 
The phase boundary between the FQ-BEC phase and SN is thus detected as the crossing of the 
$S=1$ and $S=2$ lowest excited states (see Appendix~\ref{sec:appendix_smalljj}, Fig.~\ref{fig:app_sn}(a)). 
The structure factors (Fig.~\ref{fig:app_sn}(b)) indeed indicate that 
inside the SN phase $\braket{\mathcal{Q}(\boldsymbol{k})}$ at $\Gamma$-point overwhelms 
$\mathcal{N}(\boldsymbol{k})$ at the $\Gamma$-point, 
and vise versa for the FQ-BEC-1 phase at $J'/J\lesssim -0.015$. 
The boson density in the SN phase is $\braket{n_{t}} \sim 1$ (Fig.~\ref{fig:app_sn}(d)), 
consistent with the SN phase of a spin-1 BLBQ model. 
When $B/J$ is large the lowest energy state of each dimer is a triplet, so that the bosons are fully occupied 
and the weak inter-dimer coupling works to exchange these triplets and form a SN. 
\section{Discussion}
\label{sec:discussions}
\subsection{Origin of quadurupolar moments on a dimer-bond}
\label{subsec:origin_quadrupolar}
\begin{figure}
      \centering
      \includegraphics[width=86mm]{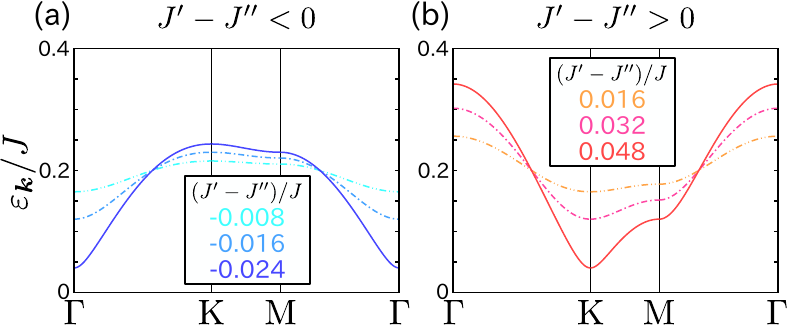}
      \caption{
                Energy bands of the eigenstates of the quadratic Hamiltonian (Eq.~(\ref{eq:boson_quadratic}))
                for (a) $J'-J'' < 0$ and (b) $J'-J'' > 0$.
                We set $B/J = 0.2$, 
                and use the parameters $t$, $P$ and $\mu$ defined in Eq.~(\ref{eq:boson_coefficients}),
                at the first order.
            }
      \label{fig:boson_energyband}
    \end{figure}

  In the FQ-BEC-1 and FQ-BEC-2 phases of our $\mathcal{S}=1$ effective model (Eq.~(\ref{eq:boson_hamiltonian})), 
  the pair-creation and annihilation term, $\mathcal{H}_{P}$ seems to play a major role as 
  indicated by the large energy gain $e_{P}$. 
  To clarify the role of $P$, we choose the parameter $J'=-J''$ to exclude the contribution from $\mathcal{J}$. 
  Then, Eq.~(\ref{eq:boson_hamiltonian}) at the first order level in $J'$ and $J''$ is reduced to 
        \begin{align}
            \mathcal{H}_{\text{quad}}
        & = - \mu \sum_{i = 1}^{N} n_{i}
                + \sum_{\braket{i, j},\alpha} 
                        \left[
                            \left( 
                                t b_{i, \alpha}^{\dag} b_{j, \alpha}
                            + P b_{i, \alpha}^{\dag} b_{j, \alpha}^{\dag}
                            \right) + \text{h.c.}
                        \right], 
            \label{eq:boson_quadratic}
        \end{align}
   which is transformed via 
  $b_{\boldsymbol{k}, \alpha}^{\dag} = \dfrac{1}{\sqrt{N}} \sum_{i = 1}^{N} b_{i, \alpha}^{\dag} \mathrm{e}^{-\mathrm{i}\boldsymbol{k}\cdot\boldsymbol{r}_{i}}$ to 
        \begin{align}
            \label{eq:boson_quadratic_FT}
            \mathcal{H}_{\text{quad}}
        & = \dfrac{1}{2} \sum_{\boldsymbol{k}, \alpha}
                \left[
                    \left( t \eta_{\boldsymbol{k}} - \mu \right)
                    \left(
                        b_{\boldsymbol{k}, \alpha}^{\dag} b_{\boldsymbol{k}, \alpha}
                    + b_{\boldsymbol{k}, \alpha} b_{\boldsymbol{k}, \alpha}^{\dag}
                    \right)
                \right. \notag \\
        &\qquad
                \left.
                + P \eta_{\boldsymbol{k}}
                    \left(
                        b_{\boldsymbol{k}, \alpha}^{\dag} b_{-\boldsymbol{k}, \alpha}^{\dag}
                    + b_{-\boldsymbol{k}, \alpha} b_{\boldsymbol{k}, \alpha}
                    \right)
                \right]
            + \text{const.}, 
        \end{align}
  where $\eta_{\boldsymbol{k}}
        = 2 \big( \cos{k_{x}} 
                + \cos{( \dfrac{k_{x} + \sqrt{3} k_{y}}{2})}
                + \cos{( \dfrac{k_{x} - \sqrt{3} k_{y}}{2} )}
                \big)$.
  Then, using the Bogoliubov transformation
        \begin{align}
            \begin{pmatrix} \beta_{\boldsymbol{k}} \\ \beta_{-\boldsymbol{k}}^{\dag} \\ \end{pmatrix}
        = \begin{pmatrix}
                \cosh{\theta} & \sinh{\theta} \\ 
                \sinh{\theta} & \cosh{\theta} \\ 
            \end{pmatrix}
            \begin{pmatrix} b_{\boldsymbol{k}} \\ b_{-\boldsymbol{k}}^{\dag} \\ \end{pmatrix}
        \end{align}
  with $\tanh{2\theta} = P\eta_{\boldsymbol{k}}/(t\eta_{\boldsymbol{k}} - \mu)$, 
  the Hamiltonian can be diagonalized as
        \begin{align}
            \mathcal{H}_{\text{quad}}
        = \sum_{\boldsymbol{k}, \alpha}
                \varepsilon_{\boldsymbol{k}}
                \left(
                    \beta_{\boldsymbol{k}, \alpha}^{\dag} \beta_{\boldsymbol{k}, \alpha} 
                + \beta_{\boldsymbol{k}, \alpha} \beta_{\boldsymbol{k}, \alpha}^{\dag}
                \right)
            + \text{const.},
        \end{align}
  where the particle-hole symmetric energy bands are obtained as 
        \begin{align}
            \varepsilon_{\boldsymbol{k}}
        = \pm \dfrac{1}{2}
                \sqrt{\left(t\eta_{\boldsymbol{k}} - \mu\right)^{2}-\left(P\eta_{\boldsymbol{k}}\right)^{2}}.
        \end{align}
  Figure~\ref{fig:boson_energyband}(a) and \ref{fig:boson_energyband}(b) show 
  $\varepsilon_{\boldsymbol{k}}/J$ for $J'-J'' < 0$ and $J'-J'' > 0$, respectively, at $B/J = 0.2$. 
  When the bottom of the band touches the zero level, 
  the instability takes place and the $\beta_{\boldsymbol{k}, \alpha}$-bosons of that wave number condense and form a BEC phase. 
  This happens by increasing $J'=-J''$ only up to $\left| J'-J'' \right| \sim 0.05$, 
  which is consistent with the numerical analysis that the singlet product state immediately gives way to the FQ phases in the $J'=-J''$ direction. 
  The wavenumber at which the $\varepsilon_{\boldsymbol{k}}$ takes the minimum is the $\Gamma$-point when $J'-J'' < 0$, 
  whereas it is the $\mathrm{K}$-point for $J'-J''>0$. 
  The former is the usual uniform FQ ordering, 
  and the latter explains well the particular three-sublattice-like structure of 
  the quadrupole correlations in FQ-BEC-2 phase we saw in Fig.~\ref{fig:correlation_circle}(b).
\subsection{Classification of ``nematic'' phases}
  Conventionally, a typical spin nematic (SN) phase in spin-1 system 
  is identified by the absense of net local-sublattice magnetic ordering and remaining quadrupolar ordering. 
  Typically in spin-1 BLBQ models written in the form, 
         $ J\boldsymbol{S}_{i} \cdot \boldsymbol{S}_{j}
        + B \left( \boldsymbol{S}_{i} \cdot \boldsymbol{S}_{j} \right)^{2}
        =  (J-B/2) \boldsymbol{S}_{i} \cdot \boldsymbol{S}_{j}
        + (B/2) \boldsymbol{Q}_{i} \cdot \boldsymbol{Q}_{j}
        + \text{const.}$, 
    the dipolar (magnetic) and quadrupolar orders compete with each other, 
    and the latter appears when $|B| \gtrsim |J|$ where the magnetic ordering is suppressed. 
    The QDJS is formed by $S=0,2,4\cdots$, indicating that the SU(2) symmetry is 
    broken by the binding of two-magnon pairs. 
\par
  Another series of spin nematics is a magnon-pair condensation in the spin-1/2 models~\cite{Zhitomirsky2010_EurophysLett_92_37001}. 
  Starting from the fully poralized ferromagnetic phase such as those in a high magnetic field, 
  two-magnon instability overwhelms the one-magnon one when there is a frustration effect that prohibits 
  the kinetic motion of a single magnon. 
  The bound pair of magnons propagate together, which is is by definition a quadrupolar order parameter itself written as 
  $\braket{b_{i\downarrow}^\dag b_{j\downarrow}^\dag}=\braket{S_{i}^{-}S_{j}^{-}} = \mathcal{Q}\mathrm{e}^{\mathrm{i} 2\theta}\neq 0$. 
  This bond-nematics is realized when the bond is almost fully occupied by bosons; 
  the adjacent bosons carrying ${\mathcal S}^z=1$ and -1 exchange or flucuate in pairs and form a quadrupole, 
  and accordingly the ${\mathcal S}^z=-1$ propage in space. 
  It is equivalent to the spin-2 BEC in cold atoms formed by $S=0$ and 2 sectors~\cite{Ueda2002_PhysRevA_65_063602}. 
  For a particular model that excludes the $S=1$ sector out of the low energy subspace, 
  this kind of spin nematics can be found at zero field, 
  which is characterized by the QDJS of bound-bosons similar to the spin-1 BLBQ case
  \cite{Shannon2006_PhysRevLett_96_027213,Momoi2012_PhysRevLett_108_057206}. 
\par
  These two types of ordering are categolized as $n$-type spin nematics, and are only possible when 
  the $\mathcal{S}=1$ bosons can be bounded by a strong quantum flucuation, 
  mostly in a strong magnetic field that helps to increase the population of $\mathcal{S}=1$ bosons. 
\par
  The other known class of spin nematic phase is the $p$-type nematic found in spin-1/2 systems~\cite{Lauchli2005_PhysRevLett_95_137206}.
  The vector chirarity of two spins $\boldsymbol{p}_{i}=\boldsymbol{S}_{i_{1}}\times\boldsymbol{S}_{i_{2}}$ defined on a bond ($\boldsymbol{p}_{i}$ in Eq.~(\ref{eq:p})) condense and form a nematic order, 
  again suppressing the sublattice magnetic ordering. 
  This time the SU(2) symmetry is broken down to U(1) of a uniaxial rotator, 
  and the QDJS consisting of $S=0,1,2,\cdots$ indicates the BEC from a one-magnon instability. 
\par
  In a 1D and 2D spin dimer systems, the quadrupolar moment is defined on a dimer-bond, and depending on the choice of 
  the parameters, one can control the number of bosons from zero to one. 
  When $\braket{n_{t}} =1$ the same situation as the first spin-1 type of SN is realized, whereas 
  for $\braket{n_{t}} <1$, a BEC type of quadrupolar ordering is realized which is FQ-BEC-1 or F-nematic. 
  In our FQ phases, a \textit{single} magnon condenses by the hopping and pair creation and annihilations, 
  breaking the SU(2) symmetry down to U(1), 
  which is an analogue of the spinor BEC in cold atoms~\cite{Ho1998_PhysRevLett_81_742}, 
  and shares the same property as the $p$-type nematics. 
  Since the dimer remains always nonmagnetic, 
  it can be regarded as some sort of ``nematic'' order in terms of $\mathcal{S}=1$ boson. 
  When separately looking at upper/lower-layer carrying small moments, they form a ferromagnetic 
  sublattice long range order, 
  whereas the intra-dimer quantum fluctuation kills the moment inside the dimer, 
  which should be regarded as a different phase from the mean-field type of 
  inter-layer AF and intra-layer ferromagnetic ordering 
  of the full spin moments observed in the ferromagnetic dimer model~\cite{Hikihara2019_PhysRevB_100_214414}. 
\par
  We saw in \S.\ref{sec:lro} that the FQ-BEC-2 phase has a strong chirality correlation 
  and the possible 120$^\circ$ in-plane magnetic long range correlation of small moments, which are masked 
  and the dimer unit remains nonmagnetic. 
  There are two possibilities; 
  In a mean-field analysis these two are incompatible, whereas our treatment treating 
  the full quantum many body effect may allow for a new possibility that the two types of order may coexist. 
  The other possibility is the absense of sublattice magnetic ordering that may stabilize the $p$-type spin nematics. 
  The Anderson tower still does not allow for the separation of the behavior of the biaxial or uniaxial rotator, 
  namely the full breaking of SU(2) or the breaking only down to U(1) 
  within the present study. 
\subsection{Exchange of \texorpdfstring{$\mathcal{S}=1$}{S=1} moments}
\label{subsec:exchange_spin1}
\begin{figure}
            \centering
            \includegraphics[width=86mm]{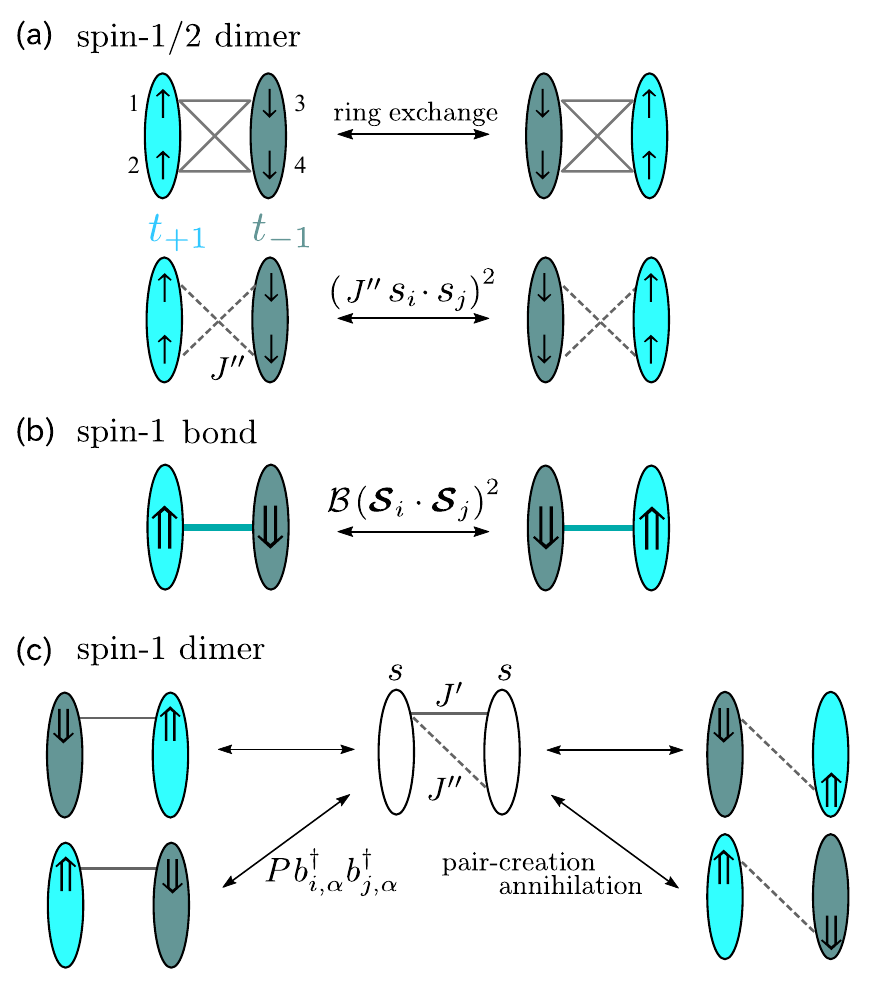}
            \caption{
                Three different types of fluctuations that contribute to the formation of the spin nematics. 
                Single and double arrows represent the spin-1/2 and spin-1, respectively. 
                (a) In spin-1/2 dimer system, ring exchange interaction that permutates spin-1/2's as 
                $(1,2,3,4) \leftrightarrow (2,3,4,1)$ in the upper panel 
                (see Refs.~\onlinecite{Yokoyama2018_PhysRevB_97_180404,Tanaka2018_JPhysSocJpn_87_023702}) 
                and the second order perturbation terms operated twice, $(J''\boldsymbol{s}_{i}\cdot \boldsymbol{s}_{j})^2$, with $\boldsymbol{s}_{i}$ the spin-1/2 operator 
                discussed in Refs.~\onlinecite{Hikihara2010_PhysRevB_81_064432,Hikihara2019_PhysRevB_100_214414}, 
                work in the similar manner. 
                (b) Fluctuation between on-bond spin-1's that are equivalent to those of panel (a). 
                (c) In our spin-1 dimer system, the pair creation and anihilation term ($P$) plays a major role 
                which originates from the first order in $J'$ and $J''$. 
            }
            \label{fig:origin}
        \end{figure}

  Previously, in a spin-1/2 dimer system~\cite{Yokoyama2018_PhysRevB_97_180404,Tanaka2018_JPhysSocJpn_87_023702}, 
  we showed that the origin of the spin nematic phase is the inter-dimer 
  ring exchange interactions that permutate the four spin-1/2 along the twisted path as, 
  $(1,2,3,4) \rightarrow (2,3,4,1)$, which is shown in Fig.~\ref{fig:origin}(a). 
  In that case, the two spin-1/2's on a dimer form an $\mathcal{S}_{i}=1$ triplet, and the ring exchange interaction 
  exchanges the spin-1's on neighboring dimers, $(\mathcal{S}_i^{z}, \mathcal{S}_{j}^{z})=(+1,-1)$ with $(-1,+1)$-states (see Fig.~\ref{fig:origin}(b)), and suppress the dipolar ordering. 
  This plays the same role as the biquadratic interaction, $\mathcal{B}(\boldsymbol{\mathcal{S}}_{i}\cdot\boldsymbol{\mathcal{S}}_{j})^{2}$, 
  and when all the dimers are filled with a triplet, $\mathcal{S}_i=1$, the system is reduced to the BLBQ model. 
\par
  In the ferromagnetically coupled spin-1/2 dimer model~\cite{Hikihara2019_PhysRevB_100_214414}, 
  it is shown that the exchange interaction, $J'$ and $J''$ ($J_{\parallel}$ and $J_{\times}$ in their notation), 
  operated twice at the second order perturbation is important to stabilize the spin nematic phase. 
  As shown schematically in Fig.~\ref{fig:origin}(a), this works in the same manner as the ring exchange interaction, 
  and generates an effective biquadratic term~\cite{Hikihara2010_PhysRevB_81_064432}. 
  However, this time they need a larger $J''$ as their spin nematics need to compete with the stable ferromagnetic phase. 
\par    
  In our spin-1 dimer, the pair-creation and annihilation term results in an off-diagonal pair condensation of up and down spin-1's 
  via the processes shown in Fig.~\ref{fig:origin}(c). 
  These processes, when performed twice, will give the same effect as the 
  biquadratic interaction in Fig.~\ref{fig:origin}(b). 
  The advantage here is that it is a first order process and can be more easily realized than $\mathcal{B}$ or the ring exchange processes.
\par
  The second order perturbation process in the spin-1/2 dimer systems giving the effective biquadratic interactions can also be understood as the pair-creation and annihilation process,
  while not discussed in the spin-1 bosonic language there~\cite{Hikihara2010_PhysRevB_81_064432,Hikihara2019_PhysRevB_100_214414}.
  Nevertheless, the pair fluctuation effect in the spin-1 dimer systems is stronger than that of spin-1/2 dimer systems; 
  One can see from Eq.~(\ref{eq:triplets}) that the triplet states $\mathcal{S}=1$ 
  consists of twice as large dimer-spin terms 
  than the triplets formed by two spin-1/2's, which means that the entanglent between the two $\mathcal{S}=1$ can be 
  easily enhanced. 
  We consider this to be the origin of a variety of quadrupolar phases in our system.
\section{Summary and perspective}
\label{sec:summary}
  In conclusion, we found various types of quadrupolar ordering in a spin-1 dimer model forming a triangular lattice bilayer. 
  In the decoupled dimer limit, the dimer-bond host singlet at small $B/J$ and triplet at larger $B/J$. 
  Including the inter-dimer Heisenberg exchange terms $J'$ and $J''$, perturbatively up to second order, 
  we derived an effective hard-core bosonic model describing the triplet on a dimer-bond, 
  which reproduces the low-energy properties of the original model. 
  The major part of the bosonic Hamiltonian consists of the hopping ($t$) and pair-creation and annihilation ($P$) of bosons, 
  as well as the chemical potential ($\mu$) and the antiferromagnetic exchange interaction ($\mathcal{J}$). 
  The bosons are doped by $\mu$ and the $t$-term contributes to the formation of BEC. 
  The ferromagnetic (FM) and antiferromagnetic (AFM) phases appear due to $\mathcal{J}<0$ and $\mathcal{J}>0$. 
  When $t$ and $P$ are dominant at $J' \sim -J''$ 
  the FQ-BEC-1 and FQ-BEC-2 phases are observed, which are the condensation of quadrupolar moments on a dimer-bond, 
  with $\sim 0.55$ bosons per dimer, similarly to the anisotropic superfluidity in cold atoms\cite{Ho1998_PhysRevLett_81_742}. 
  Also when $B/J\gtrsim 1/3$, the typical spin-1 nematic phase is found near the inter-dimer decoupling ($J', J''\sim 0$) region, 
  since each dimer is occupied by a spin-1 boson which exchange by $P$.  
\par
  Our results are widely applied to the bilayer quantum spin dimer systems, 
  since the interactions appear in Eq.~(\ref{eq:spin_hamiltonian}), are all standard ones that are derived 
  naturally from the strong coupling perturbation theory of Mott insulator; 
  the Heisenberg exchange interactions, $J$, $J'$ and $J''$, 
  and the biquadratic intra-dimer interaction $B$. 
  The value of $B$ is reported to be relatively larger than it has been believed before~\cite{Tanaka2018_JPhysSocJpn_87_023702}. 
\par
  The FQ-BEC-2 phases discussed here might be exotic in the sense that the distribution of $S=1$-bosons are uniform in space, 
  whereas the quadrupolar correlation may develop a particular spatial modulation. 
  It differs from the SN of the $S=1$ BLBQ model and from the multi-magnon-bound states of the $S=1/2$ models in high fields. 
  We consider that the magnetic long range ordering is possibly absent, and the $p$-nematic type of correlation develops toward the long range ordering. 
\par
  We finally discuss the relevance with the actual material. 
  In a family of Ba$_3M$Ru$_{2}$O$_{9}$ ($M=$ Ca, Sr, Co, Ni, Cu, Zn)~\cite{Darriet1976_JSolidStateChem_19_213,
  	Lightfoot1990_JSolidStateChem_89_174,Rijssenbeek1999_JSolidStateChem_146_65,
  	Beran2015_SolidStateSci_50_58,Terasaki2017_JPhysSocJpn_86_033702,Yamamoto2018_JPhysCondensMatter_30_355801}
  the two face-shared RuO$_{6}$ octahedra form a dimer which is stacked along the two-dimensional triangular lattice 
  in the same way as our model Fig.~\ref{fig:lattice}(a). 
  The Ru$^{5+}$ carries possibly $S=1$~\cite{Terasaki_privatecommun}
  and the material is well described by our Hamiltonian, Eq.~(\ref{eq:spin_hamiltonian}). 
  Intriguing magnetic properties were reported; 
  for a Zn-compound, the uniform susceptibility is strongly suppressed down to 37~mK, 
  a much lower temperature than the value of $J\sim 150-200$~K~\cite{Terasaki2017_JPhysSocJpn_86_033702,Yamamoto2018_JPhysCondensMatter_30_355801}. 
  In a Co-, Ni- or Cu-compound with shorter inter-dimer distances, namely having larger $|J'/J|$ and $|J''/J|$, 
  the system undergoes an antiferromagnetic transition at $T_{\mathrm{N}}\sim 100$~K~\cite{Lightfoot1990_JSolidStateChem_89_174,Rijssenbeek1999_JSolidStateChem_146_65}. 
  The Ca- and Sr-compound with the longer inter-dimer distances contrarily favor a standard nonmagnetic singlet state~\cite{Darriet1976_JSolidStateChem_19_213}. 
  Usually, $J$, $J'$ and $J''$ are antiferromagnetic ones, and if we increase the inter-dimer interactions 
  from the center of the phase diagram in Eq.~(\ref{fig:phasediagram_2ndorder}) toward the upper right direction, 
  the ground state transforms from singlet, FQ-BEC-2, and to an antiferromagnetic phase, in good agreement with the 
  experimental observation, (Sr, Ca)$\rightarrow$(Zn)$\rightarrow$(Co, Ni, Cu). 
  Since the conventional SN phases were all found next to the fully polarized ferromagnetic/antiferromagnetic phase, 
  the exotic nonmagnetic phase in the Zn-compound was not really connected to the quadrupolar ordering. 
  Our series of studies on spin dimer systems~\cite{Yokoyama2018_PhysRevB_97_180404} 
  point out that SN and other quadrupolar phases of spin-1 can be found next to the spin-0 singlet phase. 
  Since this kind of symmetry breaking is not directly detected from the standard susceptibility measurements, 
  the way to identify them experimentally should be discussed in the next step for the 
  clarification of the unexplored nature of the spin dimer materials. 
\begin{acknowledgments}
  We thank Keisuke Totsuka, Tsutomu Momoi and Toshiya Hikihara for helpful discussions,
  Yuto Yokoyama for advice in the early stage of this work, 
  Masataka Kawano for helpful comments, and Ichiro Terasaki for fruitful communications.
  This work is supported by JSPS KAKENHI Grants No. JP17K05533, No. JP18H01173, No. JP17K05497, 
  No. JP17H02916. 
\end{acknowledgments}
\appendix
\section{Details on the effective model Eq.~(\ref{eq:boson_hamiltonian})}
\label{sec:appendix_effectivemodel}
\subsection{Construction of a time-reversal invariant basis of spin-1 dimer state via bond-operator approach}
\label{subsec:appendix_bondoperator}
  To make direct connections of our representation in the main text with the previous studies, 
  we introduce the bond-operator representation of the spin-1 state in a unit of dimer.
  Previous bond-operator representations broke the time-reversal symmetry~\cite{Brenig2001_PhysRevB_64_214413,Wang2000_PhysRevB_61_4019,Kumar2010_PhysRevB_82_054404},
  which we modify to the one that keeps the time-reversal symmetry. 
\par
  First, we write down the time-reversal invariant multiplet states of spin-1 dimers.
  Using the time-reversal invariant basis for a single spin-1 state~\cite{Lauchli2006_PhysRevLett_97_087205,
    	Penc2011_IFM_Chap13_SpinNematics}
        \begin{align}
            \ket{x} = \dfrac{\mathrm{i}\left( \ket{+1} - \ket{-1} \right)}{\sqrt{2}}, \
            \ket{y} = \dfrac{\ket{+1} + \ket{-1}}{\sqrt{2}}, \
            \ket{z} = - \mathrm{i} \ket{0},
        \end{align}
  the singlet, triplet, and quintet dimer states can be rewritten as 
        \begin{align}
        \ket{s} & = \dfrac{1}{\sqrt{3}} \left( \ket{x, x} + \ket{y, y} + \ket{z, z} \right), \\
        \ket{t_{\alpha}} & = - \dfrac{1}{\sqrt{2}} \sum_{\beta, \gamma} \varepsilon_{\alpha\beta\gamma} \ket{\beta, \gamma}, \\
        \ket{q_{\alpha\beta}} & = - \dfrac{1}{\sqrt{2}} \left( \ket{\alpha, \beta} + \ket{\beta, \alpha} \right)+ \left( \sqrt{2} - 1 \right) \delta_{\alpha\beta} \ket{\alpha, \alpha},
        \end{align}
  respectively ($\alpha, \beta = x, y, z$).
  Since only three of four states $\ket{s}$, $\ket{q_{\alpha\alpha}}$ are linearly independent, 
  we construct two quintet states, $\ket{q_{3\alpha^{2}-r^{2}}}$ and $\ket{q_{\beta^{2}-\gamma^{2}}}$,
  as the linear combinations of $\ket{q_{\alpha\alpha}}$,
  whose forms are, for example,
        \begin{align}
            \ket{q_{3z^{2}-r^{2}}} 
        & = \dfrac{2\ket{q_{zz}}-\ket{q_{xx}}-\ket{q_{yy}}}{\sqrt{6}}
          = - \dfrac{2 \ket{z, z} - \ket{x, x} - \ket{y, y}}{\sqrt{6}}, \notag \\
            \ket{q_{x^{2}-y^{2}}} 
        & = \dfrac{\ket{q_{xx}}-\ket{q_{yy}}}{\sqrt{2}}
          = - \dfrac{\ket{x, x} - \ket{y, y}}{\sqrt{2}},
        \end{align}
  where we take $(\alpha, \beta, \gamma) = (z, x, y)$.
  We use the $\ket{s}$ and $\ket{q_{3\alpha^{2}-r^{2}}}$ and $\ket{q_{\beta^{2}-\gamma^{2}}}$ as a basis for representing $S_{i_{\mu}}^{\alpha}$,
  namely,
  the choice of the basis $\ket{q_{3\alpha^{2}-r^{2}}}$ and $\ket{q_{\beta^{2}-\gamma^{2}}}$ is dependent on $\alpha = x, y, z$.
\par
  In the main text, we adopted the singlet state as a vacuum, 
  whereas in this bond-operator approach, we redefine the vacuum as the state without any multiplet. 
  Accordingly, instead of $b_{i, \alpha}$ and $b_{i, \alpha}^{\dag}$, 
  we use $s_{i}$ ($s_{i}^{\dag}$) and $t_{i, \alpha}$ ($t_{i, \alpha}^{\dag}$) as 
  the annihilation (creation) operators of the singlet and triplet of compopnent-$\alpha$,
  and $q_{i, \alpha}$ ($q_{i, \alpha}^{\dag}$) as the ones of the quintet of component-$\alpha$.
  Then, the spin-1 operator $S_{i_{\mu}}^{\alpha}$ ($\mu = 1, 2$) in the $i$-th dimer can be written as follows;
        \begin{align}
            S_{i_{1}}^{\alpha}
        & = \mathrm{i}\dfrac{\sqrt{2}}{\sqrt{3}}
                    \left( t_{i, \alpha}^{\dag} s_{i} - s_{i}^{\dag} t_{i, \alpha} \right)
            - \dfrac{\mathrm{i}}{2} \sum_{\beta, \gamma}
                    \varepsilon_{\alpha\beta\gamma} t_{i, \beta}^{\dag} t_{i, \gamma} \notag \\
        &\quad
            - \dfrac{\mathrm{i}}{\sqrt{3}}
                    \left( q_{i, 3\alpha^{2}-r^{2}}^{\dag} t_{i, \alpha} - t_{i, \alpha}^{\dag} q_{i, 3\alpha^{2}-r^{2}} \right) \notag \\
        &\quad
            - \dfrac{\mathrm{i}}{2} \sum_{\beta \neq \alpha}
                    \left( q_{i, \alpha\beta}^{\dag} t_{i, \beta} - t_{i, \beta}^{\dag} q_{i, \alpha\beta} \right)
            - \dfrac{\mathrm{i}}{2} \sum_{\beta, \gamma} 
                    \varepsilon_{\alpha\beta\gamma} q_{i, \alpha\beta}^{\dag} q_{i, \gamma\alpha} \notag \\
        &\quad
            - \dfrac{\mathrm{i}}{2} \sum_{\beta, \gamma}
                    \varepsilon_{\alpha\beta\gamma}
                        \left( q_{i, \beta^{2}-\gamma^{2}}^{\dag} q_{i, \beta\gamma} - q_{i, \beta\gamma}^{\dag} q_{i, \beta^{2}-\gamma^{2}} \right), \notag \\
            S_{i_{2}}^{\alpha}
        & = - \mathrm{i}\dfrac{\sqrt{2}}{\sqrt{3}}
                    \left( t_{i, \alpha}^{\dag} s_{i} - s_{i}^{\dag} t_{i, \alpha} \right)
            - \dfrac{\mathrm{i}}{2} \sum_{\beta, \gamma}
                    \varepsilon_{\alpha\beta\gamma} t_{i, \beta}^{\dag} t_{i, \gamma} \notag \\
        &\quad
            + \dfrac{\mathrm{i}}{\sqrt{3}}
                    \left( q_{i, 3\alpha^{2}-r^{2}}^{\dag} t_{i, \alpha} - t_{i, \alpha}^{\dag} q_{i, 3\alpha^{2}-r^{2}} \right) \notag \\
        &\quad
            + \dfrac{\mathrm{i}}{2} \sum_{\beta \neq \alpha}
                    \left( q_{i, \alpha\beta}^{\dag} t_{i, \beta} - t_{i, \beta}^{\dag} q_{i, \alpha\beta} \right)
            - \dfrac{\mathrm{i}}{2} \sum_{\beta, \gamma} 
                    \varepsilon_{\alpha\beta\gamma} q_{i, \alpha\beta}^{\dag} q_{i, \gamma\alpha} \notag \\
        &\quad
            - \dfrac{\mathrm{i}}{2} \sum_{\beta, \gamma}
                    \varepsilon_{\alpha\beta\gamma}
                    \left( q_{i, \beta^{2}-\gamma^{2}}^{\dag} q_{i, \beta\gamma} - q_{i, \beta\gamma}^{\dag} q_{i, \beta^{2}-\gamma^{2}} \right).
        \end{align}
\subsection{Evaluation of the effective model}
\label{subsec:appendix_evaluation}
  We examine the effect of the three-dimer interactions $\mathcal{H}_{\text{3body}}$ 
  in the effective Hamiltonian $\mathcal{H}_{\text{eff}}$ (Eq.~(\ref{eq:boson_hamiltonian})), 
  which was discarded in the calculation in main text.
\par
\begin{figure}
            \centering
            \includegraphics[width=86mm]{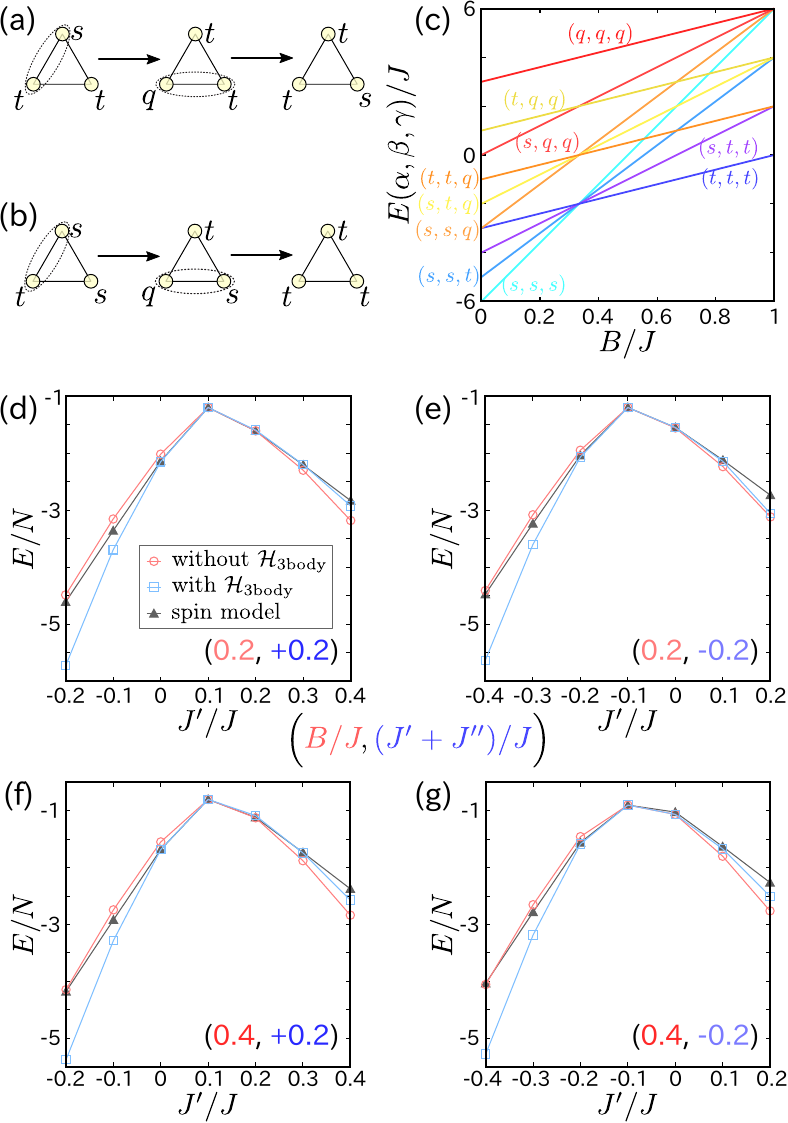}
            \caption{
                (a), (b) Typical second order perturbation processes over three-dimers.
                (a) Processes of ``correlated hopping'' and (b) pair-creation of bosons.
                Ellipses mark the pair of sites to which the perturbation Hamiltonian 
                $\mathcal{H}_{\text{inter}}$ operates. 
                (c) Energy levels of $\mathcal{H}_{\text{intra}}$ of the three spin-1 dimer states. 
                (d)--(g) $J'/J$ dependences of the ground state energies of the effective Hamiltonian 
                $\mathcal{H}_{\text{eff}}$ with and without $\mathcal{H}_{\text{3body}}$ 
                and the original spin Hamiltonian $\mathcal{H}$ on the 9-dimer triangular lattice 
                for $\left( B/J, (J'+J'')/J \right) = $
                (d) $(0.2, +0.2)$, (e) $(0.2, -0.2)$, (f) $(0.4, +0.2)$, and (g) $(0.4, -0.2)$.
            }
            \label{fig:3dimerinteraction}
        \end{figure}

  First, we show some details of $\mathcal{H}_{\text{3body}}$, 
  which originates from the three-dimer processes at the second order of perturbation. 
  We show two examples of these processes in Figs.~\ref{fig:3dimerinteraction}(a) and \ref{fig:3dimerinteraction}(b),
  where $s$, $t$, $q$ are the singlet, triplet, and quintet states, respectively, on a dimer.
  Figure~\ref{fig:3dimerinteraction}(a) is the processes similar to the correlated hoppings 
  found in the Shastry--Sutherland model~\cite{Momoi2000_PhysRevB_61_3231,Momoi2000_PhysRevB_62_15067},
  and Fig.~\ref{fig:3dimerinteraction}(b) is the pair-creation of bosons.
\par
  In treating these three-dimer processes, 
  we examined the validity of restricting the low-energy manifold of states to those including only singlet and triplets.
  Figure~\ref{fig:3dimerinteraction}(c) shows the energy diagram of the three-dimer states, 
  $E(\alpha, \beta, \gamma)$ ($\alpha, \beta, \gamma = s, t, q$). 
  We see that $(t, t, t)$ states and $(s, s, q)$ states are degenerate at $B/J = 0$,
  whereas they are well separated when a small $B/J > 0$ is introduced. 
\par
  Next, we compare the ground state energies of the effective model $\mathcal{H}_{\text{eff}}$ (Eq.~(\ref{eq:boson_hamiltonian})) 
  with and without $\mathcal{H}_{\text{3body}}$, 
  and the energy of the original spin-1 dimer model $\mathcal{H}$ (Eq.~(\ref{eq:spin_hamiltonian})). 
  We used the 9-dimer triangular lattice under the periodic boundary condition. 
  The cases of $B/J = 0.2$ are shown in Figs.~\ref{fig:3dimerinteraction}(d) and \ref{fig:3dimerinteraction}(e), 
  and those of $B/J = 0.4$ are in Figs.~\ref{fig:3dimerinteraction}(f) and \ref{fig:3dimerinteraction}(g), 
  where the parameters are chosen as $(J'+J'')/J = +0.2$ and $-0.2$. 
  It is confirmed that the energies of $\mathcal{H}_{\text{eff}}$ with $\mathcal{H}_{\text{3dimer}}$ are not always closer to those of $\mathcal{H}$ 
  than those of $\mathcal{H}_{\text{eff}}$ without $\mathcal{H}_{\text{3dimer}}$ 
  although $\mathcal{H}_{\text{eff}}$ with $\mathcal{H}_{\text{3dimer}}$ fully takes the second order perturbation terms into account. 
  We see that for $|J'/J|$ and $|J''/J| \lesssim 0.2$, the energies are in good consistency with each other. 
  The effective model may not hold quantitatively when either of $J'$, $J''$ has a large value. 
  This might be because the three-dimer interactions appear at the higher order of $J'$ and $J''$, 
  which would cancel out the three-dimer interactions derived at the 2nd order.  
\par
  As we already saw in \S.~\ref{subsec:perturbation_order}, the effect of second order perturbation is small, 
  and setting $\mathcal{H}_{\text{3body}}=0$ does not change both the quantitative and qualitative aspects of the results. 
  The advantage of having a simple Hamiltonian, $\mathcal{H}_{\text{eff}}$, is that it corresponds exactly to the spin-1/2 model, 
  and resultantly, the two models of different spin numbers can be compared on equal ground. 
\section{The small \texorpdfstring{$J'/J$}{J'/J}, \texorpdfstring{$J''/J$}{J''/J} region of the phase diagram}
\label{sec:appendix_smalljj}
\begin{figure}
      \centering
      \includegraphics[width=86mm]{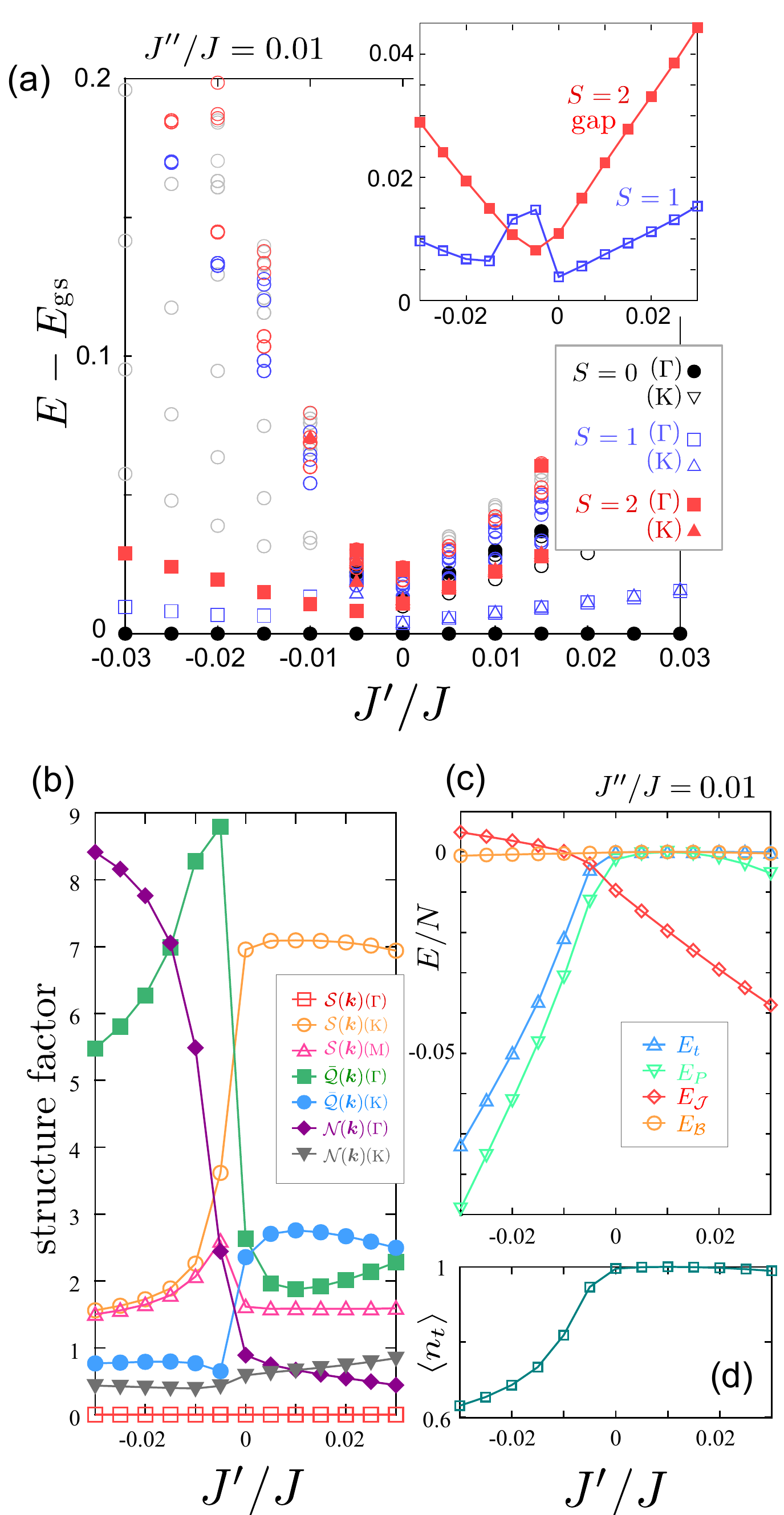}
      \caption{(a) Low-energy excited states of $S=0,1,2$ sectors 
                as a function of small $J'/J$ with fixed $J''/J=0.01$, 
                where we find a crossing of the low lying $S=1$ and $S=2$ 
                levels shown more clearly in the inset (extracted data from the main panel). 
                The region near $J'\sim 0$ where $S=2$ is the lowest excited state 
                is the typical spin nematic phase found in the $S=1$ BLBQ models. 
                (b)  Spin (Eq.~(\ref{eq:spin_structure_factor})), quadrupole (Eq.~(\ref{eq:quadrupole_structure_factor})) , and
                staggered spin (Eq.~(\ref{eq:staggered_spin_structurefactor}))
                structure factors at $\Gamma$, $\mathrm{K}$, $\mathrm{M}$-points. 
 		            (c) The contributions from major terms in the effective Hamiltonian, 
                $e_{t}, e_{P}, e_{\mathcal{J}}$ and $e_{\mathcal{B}}$. 
                (d) Triplet densities $\braket{n_{t}}$. 
            }
      \label{fig:app_sn}
    \end{figure}

	We show the details of the physical quantities in the small $J'/J$, $J''/J$ region at $B/J = 0.4$.
	Figure~\ref{fig:app_sn}(a) shows the $J'/J$ dependece of the low-energy excited states with $J''/J = 0.01$,
	together with the spin gaps of $\Delta S = 1$ and $\Delta S = 2$ in the inset.
	In varying $J'$, the $S = 1$ and $S = 2$ excited states cross at around $J'/J = -0.01$, 
	which indicates the quantum phase transition from the FQ-BEC-1 phase to the SN phase.
	The phase transition is observed in the changes of the structure factors 
        shown in Fig.~\ref{fig:app_sn}(b), 
	where $\mathcal{N}(\boldsymbol{k})$ at $\Gamma$-point is supressed at $J'/J \sim -0.01$,
	and $\mathcal{Q}(\boldsymbol{k})$ at $\Gamma$-point solely develops.
\par
	When $J'$ is increased further,
	the $S = 2$ lowest excited state again becomes higher in energy than the $S = 1$ excited state, 
	which signals the transition from the SN phase to the AFM phase~\ref{fig:app_sn}(a).
	There, $\mathcal{Q}(\boldsymbol{k})$ becomes smaller,
	and $\mathcal{S}(\boldsymbol{k})$ at K-point develops.
\par
	Figure~\ref{fig:app_sn}(c) shows the contributions of some terms in the effective Hamiltonian (Eq.~(\ref{eq:boson_hamiltonian})) to the 
        ground state energy. 
	In the FQ-BEC-1 phase, hoppings ($e_{t}$), and pair-creation and annihilation ($e_{P}$) 
        support the FQ-BEC-1 phase, as discussed in \S\S.~\ref{subsec:origin_quadrupolar}.
	In the SN phase at $-0.01 \lesssim J'/J \lesssim 0$, 
	the energy gain of $e_{P}$ is still dominant. 
	The effective biquadratic interactions generated by 
        operating pair-creation and annihilation terms twice (see~\S\S.~\ref{subsec:exchange_spin1}) 
        plays a key role, while the energy gain from the biquadratic interaction term in the Hamiltonian 
        $e_{\mathcal{B}}$ is small. 
	The AFM phase is stabilized by the gain $e_{\mathcal{J}}$ from 
        the Heisenberg exchange between the $\mathcal{S}=1$ bosons. 
\par
	The $J'/J$ dependence of the triplet density $\braket{n_{t}}$ is plotted in Fig.~\ref{fig:app_sn}(d). 
	In the vicinity of the phase transition from FQ-BEC-1 to SN,
	the triplet density rapidly increases to $\braket{n_{t}} \sim 1$.
%
\end{document}